\documentclass[pdflatex,sn-mathphys]{sn-jnl}
\usepackage{breqn}

\jyear{2021}%

\theoremstyle{thmstyleone}%
\theoremstyle{thmstyletwo}%

\theoremstyle{thmstylethree}%

\begin{document}

\title[Article Title]{Investigation of Dust Ion Acoustic Shock and Solitary Waves in a Viscous Dusty Plasma}


\author{\fnm{J.} \sur{Goswami}}

\author*{\fnm{S. S.} \sur{Kausik*}}\email{kausikss@rediffmail.com}

\affil{\orgname{Centre of Plasma Physics, Institute for Plasma Research}, \orgaddress{\street{Nazirakhat}, \city{Sonapur-782402}, \postcode{Kamrup (M)}, \state{Assam}, \country{India}}}


\abstract{A viscous dusty plasma containing Kappa-$(\kappa -)$ distributed electrons, positive warm viscous ions and constant negatively charged dust grains with viscosity have been considered to study the modes of dust-ion-acoustic waves (DIAWs) theoretically and numerically. The derivations and basic features of shock and solitary waves with different plasma parameters like Mach number, finite temperature coefficient, unperturbed dust streaming velocity, kinematic viscosity of dust etc. of this DIAWs mode have been performed. Considering the dynamical equation from Korteweg–de Vries(KdV) equation, a phase portrait has been drawn and the position of saddle point or col. and center have also been discussed.  This type of dusty plasma can be found in celestial bodies. The results of this research work can be applied to study the properties of DIAWs in various astrophysical situation where $\kappa$-distributive electrons are present and careful modification of the same model can help us to understand the nature of the DIAWs of laboratory plasma as well.}




\maketitle

\section{Introduction}\label{sec1}
In the past few years, dusty plasmas or complex plasmas have become a topic of great research interest around the globe. It is a kind of plasma where nanometer or micrometer-sized particles are suspended. In most of the cases, these nanometer or micrometer sized particles known as dust are negatively charged but there are various circumstances where they may be positively charged. The charge of dust particles varied in its size, greater the size of the dust particles larger its charge will be. Coupling via electrostatic force or any other forces like collision or viscous may happen and this coupling can change from weakly coupled to crystallising structures in the presence of these dust particles. The dependence of coupling parameter with the structural parameter has been discussed in detail by Dragan and Kutarov \cite{Dragan}. As we consider the density of dust particles is low, the viscosity is also very low, and no collision is being considered, so the effect of coupling is not taken into account in this present model like many other profound scientists did for dust ion acoustics waves (DIAWs)\cite{Patrice,patrice2017nonlinear}. Dusty plasma can be found in the celestial objects like comets, zodiacal dust cloud, interstellar, circumstellar clouds etc \cite{shukla2015introduction}. Not only astrophysical bodies but there are many examples of laboratory dusty plasma as well \cite{shukla2015introduction}.
\par
Depending on the motion of the species, dusty plasma can produce various waves, such as dust-acoustic waves (DAWs) \cite{rao1990dust}, dust ion acoustic waves (DIAWs) \cite{shukla1992dust}. We are going to consider dust ion acoustic waves (DIAWs) as the motion of both ions and dust species and also the distribution of electrons are taken into account. Some authors have mentioned the dust ion-acoustic waves term only by considering the motion of ions \cite{d1994ion,bansal2020nonplanar,ma1994self,vranjevs2002ion,MAmundust,HASSAN2022105106,Roy2014,ADHIKARY20121460}.
\par
Furthermore, the ion-acoustic (IA) \cite{goswami2018study}, electron acoustic (EA) \cite{devanandhan2011electron} and electron-static (ES) \cite{goswami2020collision} modes can also be present in a dusty plasma environment.
\par
Also, shock waves in plasma have been found to have a greater interest in recent years. Various researchers have found many interesting results in the formation of shock waves in IA, EA, ES cases, and Laser Plasma Interactions (LPI) too \cite{ghosh2021resonant}. Both theoretical and experimental study of shock waves has been found in recent years \cite{barkan1996experiments,nakamura1999observation,Paul_2022,Nakamura2}. In the case of the production of shock waves in dusty plasma, there exists a good number of theoretical and experimental works \cite{Nakamura2,KaurSaini+2018+215+223,Haider,denra2018study}. Shock waves can be produced in plasma if there is a dissipative force present in the plasma. Two very important dissipative forces arise in the plasma due to collision and viscosity. These two forces cause the damping of plasma because of which shock waves have been formed.
\par
In low density astrophysical plasma domain, there is deviation of the velocity distribution function (VDF) of particle from the Maxwellian distribution due to less number of binary collision among charges. Lorentzian velocity distribution function describes such superthermal particle distribution. Kappa distribution is one of these kinds of superthermal distribution where the particles have high-energy tails. This distribution can be given as\cite{pierrard2010kappa,goswami2021amplitude}-
\begin{equation*}
    f_{\kappa}^j (v) = \frac{n_{j0}\Gamma(\kappa + 1)}{\left(\pi \kappa \theta^2 \right)^{3/2} \Gamma(\kappa-\frac{1}{2})}\left[1+\frac{v^2}{\kappa \theta^2}\right]^{-(\kappa+1)}.
\end{equation*}
Here, $n_{j0}$ is the unperturbed particle density and $\theta^2 = \frac{\left(\kappa-\frac{3}{2}\right)}{\kappa}\frac{2k_B T_j}{m_j}$. The effect of $\kappa$-distributive electrons have been shown by various researchers \cite{ATTEYA20181931,baluku2008dust,liu2009dust}. In the present paper, we have considered collisionless space plasma, and that is why we have taken kappa-distributed electrons. This type of distribution is frequently presented in collisionless space plasma \cite{Nicolaou_2018}.
\par
At the time of the formation of the model of this present work, we have kept mainly two types of space plasma in our minds.
\par
(a) Dusty plasma in an interstellar molecular cloud, a dense plasma, can consist of an H+ ion. In these types of dusty plasma, the normalized charge density of the dust grains is one (q=1).\cite{EL-Labany1,Shan1}.
\par
and
\par
(b) Dusty plasma is produced by the reaction of pickup $H^+$ ions and ambient plasma (i.e., solar wind) \cite{MANN2008160,schwadron2007pickup,Holzer}. 
\par
When physical parameters are varied there is a qualitative change in the behaviour of the system, these changes will be found in the Bifurcation analysis. Dynamical behaviours of nonlinear DIAWs in a magnetized dusty plasma has been reported by Samanta \emph{et al.} \cite{Samanta}.
\par
In this work, the motion of ion and dust grains have been considered to construct this model and in both the cases kinematic viscosity is present. Also a finite temperature correction for the ions has also been taken in this model. As the time scale of the electrons is very much less than the other two species so we have taken a distribution for the electrons which is the kappa distribution. The work is performed in the weak non-linearity region and that is why we have used reductive perturbation technique (RPT). The momentum in the system are majorly supplied by the dust grains as it has heavier mass than ions and on the other hand the force in the system is majorly supplied by the ions. So keeping these things in mind we have derived the KdV-Burger and KdV equations for the system and most importantly discussed about the weak and strong shock in a DIAWs mode for the first time.
\par
This paper is organised in the following way- in section \ref{sec2} we have formulated the problem for the dusty plasma with the detailed discussion on kinematic viscosity and finite temperature correction in the ion in the following two subsections \ref{sec2A} and \ref{sec2B}. In \ref{sec3} we have linearized the dynamic equations given in \ref{sec2} and achieved the linear dispersion relation for real and imaginary cases, as two dispersion relation arise here have the same form that is why we have discussed a general solution for both the equations in the subsection \ref{sec3A}. In the next section \ref{sec4} we have derived the KdV-Burger(KdVB) equation and in the subsection \ref{sec4A} we have solved the same.
Afterward, under subsection \ref{sec4B}, without considering the viscosity of both the species (dust grains and the ions), we have derived the KdV equation. The analytical and the numerical solution of the KdV equation have been achieved in subsections \ref{sec4B} and \ref{sec4C}. Next, we have discussed the dynamic character and the stability of the KdVB equation in the section \ref{sec5}. Then we have discussed the results and discussion of the solution of KdVB and KdV equation which eventually giving us the shock fronts and solitary structures in this section \ref{sec6}. Finally, in section \ref{sec7} we give a brief summary of this paper.
\section{Basic Formulations:}\label{sec2}
We have considered a unmagnetized dusty plasma in astrophysical space and investigated the dust-ion acoustic modes in such a plasma. As we have investigated the DIA mode, the dynamical equations for both ion and negative dust have been taken into account. The continuity equations for both the species is given by:
\begin{equation}\label{eq1}
    \frac{\partial n_j}{\partial t}+\frac{\partial \left( n_j u_j \right)}{\partial x}=0 ,
\end{equation}
here $j=d,i$. Now, the normalized momentum equation for the negative dust species in presence of the viscosity is given by
\begin{equation}\label{eq2}
    \frac{\partial u_d}{\partial t}+u_d \frac{\partial u_d}{\partial x}=q \frac{\partial \phi}{\partial x}+\eta_d\frac{\partial^2 u_d}{\partial x^2}.
\end{equation}
The momentum equations for the ions is given by
\begin{equation}\label{eq3}
    \frac{\partial u_i}{\partial t}+u_i \frac{\partial u_i}{\partial x}=-\frac{\partial \phi}{\partial x}-3\sigma n_i \frac{\partial n_i}{\partial x}-\eta_i \frac{\partial^2 u_i}{\partial x^2}
\end{equation}
These continuity and momentum equations are bounded by the Poisson's equation which is given by,
\begin{equation}\label{eq4}
    \frac{\partial^2 \phi}{\partial x^2}=n_e-\delta n_i+q n_d
\end{equation}
The electron has a super thermal distribution here in equation \ref{eq4} and the electron density can be obtained after normalising and integrating the distribution over velocity space as,
\begin{equation}\label{eq4a}
    n_e = n_{e0}\left\{1-\frac{\phi}{(\kappa - \frac{3}{2})}\right\}^{-\kappa+\frac{1}{2}} .
\end{equation}
All the other quantities have been discussed in the next three subsections \ref{sec2A}-\ref{sec2C}.
\subsection{The kinematic viscosity:}\label{sec2A}
In the momentum equations for the negative dust and positive ion, we have considered that an effect of kinematic viscosity is present. It is a very special kind of viscosity depending predominately on the velocity of that species. Coefficient of absolute viscosity $(\mu)$ is basically the measurement of internal resistance. A force is generated due to this internal resistance and is given by $F=\mu A \times S$, where $A$ surface area of the plasma and $S$ is shear rate in the plasma. This law of the force is known as ``Newton's law of viscosity''. In the equations \ref{eq2} and \ref{eq3}, the viscosity we use is the kinematic viscosity as we mention earlier and this is basically the ratio between absolute viscosity $(\mu)$ and fluid mass density (n),  this is previously discussed by Goswami and Sarkar \cite{goswami2021kbm}. 
\subsection{The finite temperature correction in the ion:}\label{sec2B}
The second term in the right hand side of the momentum equation of ion  (equation \ref{eq3}) arises due to the finite temperature correction \cite{misra2007saddle,Sen2008}. This finite temperature correction in the system only arises when the ions are going to follow  the Fermi-Dirac (FD) statics. As ions are very much lighter than the dust particle then the quantum mechanical tunneling effect comes into act and we have to consider that the chemical potential $(\mu_c)$ remains constant due to non-equilibrium dynamics. Considering all these, the non-equilibrium particle density is given by the following equation:
\begin{equation*}
    n_0=\frac{1}{2\pi^2}\left(\frac{2m}{\hbar}\right)^2 \int_{0}^{\infty}\frac{E^{1/2} dE}{e^{\beta(E-\mu_c)}+1}
\end{equation*}
where $m$ is the mass of the ion, $\hbar$ is the reduced Planck constant, $n_0$ is the equilibrium number density, $\beta=\frac{1}{k_B T_{i0}}$, $T_{i0}$ is the back-ground temperature of ion and $\mu_c$ is the chemical potential. Applying this pressure term without considering the Bohm diffraction term we get the equation \ref{eq3}. Following the same procedure as used by Akbari-Moghanjoughi and Eliasson \cite{Akbari-Moghanjoughi2016} and Goswami \emph{et al.} \cite{goswami2020collision} we get the $\sigma$ as
\begin{equation*}
    \sigma=\frac{L_{5/2}\left(-e^{\beta \mu_c}\right)}{L_{3/2}\left(-e^{\beta \mu_c}\right)}\left(\frac{V_{Ti}}{V_{Fi}}\right)^2.
\end{equation*}
Here $Li_\nu(x)$ is the polylogarithmic function in $x$ of the order $\nu$.
\par
Some of the authors \cite{lifshitz2013statistical,Masood} consider the plasma as a Fermi gas so the pressure is:
\begin{equation*}
    p_i=\frac{1}{3}\frac{m v_{Fi}^2}{n_{i0}^2}n_{i} ^3
\end{equation*}
And in this model if we consider this pressure then it will only be present in case of the ion as it is lighter than the dust species. Considering this pressure, our momentum equation \ref{eq3} would look like,
\begin{equation*}
    \frac{\partial u_i}{\partial t}+u_i \frac{\partial u_i}{\partial x}=-\frac{\partial \phi}{\partial x}-\sigma_i n_i \frac{\partial n_i}{\partial x}-\eta_i \frac{\partial^2 u_i}{\partial x^2}
\end{equation*}
where $\sigma_i$ is the ratio between ion to electron Fermi temperature.
\subsection{Normalization Schemes and Charge Neutrality}\label{sec2C}
From equations \ref{eq1}-\ref{eq4}, all the equations are normalized and the standard normalization schemes are given by: $\hat{x} \xrightarrow[]{} \frac{x}{\lambda_D}$, $\hat{t} \xrightarrow[]{} \omega_p t$, $\hat{\phi} \xrightarrow[]{} \frac{e \phi}{T_i}$, $\hat{u_j} \xrightarrow[]{} \frac{u_j}{v_{Te}}$, and $\hat{\eta_{j}} \xrightarrow[]{} \frac{\eta_j}{\lambda_D v_{Te}}$. Also, at the equilibrium the charge neutrality condition is $n_{e0}+qn_{d0}=\delta n_{i0}$. Generally, plasma parameters have effects on the function of dust charge. But dust-ion acoustic (DIA) time scale is much shorter than the dust charging time scale. Thus, we can clearly assumed that in DIA mode there is no significant effect of dust charge fluctuation. Therefore, the dust charge can be assumed to be constant \cite{el2015kinematic,shahmansouri2013dissipative}. In this manuscript we have all along taken `$q=1$'. The various values of this dust charges in different astrophysical plasmas have been discussed by EL-Labany \emph{et al.} \cite{EL-Labany1}. 
\section{Linear Dispersion Relation:}\label{sec3}
In order to investigate the linear and nonlinear behavior of dust acoustic wave in this three-component electron–ion-dust plasma,we make the following perturbation expansion for the field quantities $n_j$, $u_j$ and $\phi$ about their equilibrium values: 
\begin{equation}\label{eq5}
\begin{bmatrix} n_{j} \\ u_{j} \\ \phi \end{bmatrix}=\begin{bmatrix} 1 \\ u_{j0} \\ \phi_0 \end{bmatrix}+\varepsilon \begin{bmatrix} {n_j}^{(1)} \\ {u_j}^{(1)} \\ \phi^{(1)} \end{bmatrix}+\varepsilon^2 \begin{bmatrix} {n_j}^{(2)} \\ {u_j}^{(2)} \\ \phi^{(2)} \end{bmatrix}+...
\end{equation}
We assumed that all field variables varying as $exp[i(k x - \omega t)]$ and here $\omega$ is the normalized wave frequency and $k$ is the wave number, which contains both real$(k_r)$ and imaginary part$(k_i)$. Here, the viscous term plays a very pivotal role. The dispersion equation has an exponentially decaying complex part in addition to the real dispersion relation. In this case if we substitute the wave number with a real plus imaginary parts(given by $k = k_r + ik_i$), we obtain the two dispersion relations \textbf{as} given below. 
\par
The real dispersion relation is given by,
\begin{equation}\label{eq6}
    P_1 {\omega'}^4+Q_1 {\omega'}^3+R_1{\omega'}^2+S_1 {\omega'}+T_1=0,
\end{equation}
where,
\begin{equation}\label{eq7}
\left. \begin{array}{ll}
\displaystyle P_1=-\left[{k_r}^2+\frac{1-2\kappa}{2\kappa-3}\right]
  \\[8pt]
  \displaystyle Q_1=-2 k_r P_1\left(u_{i0}+u_{d0}\right)\\[8pt]
  \displaystyle R_1=P_1 \left[4{k_r}^2 u_{i0} u_{d0}+{k_r}^2 {u_{d0}}^2+{k_r}^2 {u_{i0}}^2 +3 \sigma {k_r}^2\right]+2{k_r}^2 \left(\delta+q^2\right) \\[8pt]
  \displaystyle S_1=-P\left(2{k_r}^3 u_{i0} {u_{d0}}^2+2{k_r}^3 {u_{i0}}^2 {u_{d0}}+6{k_r}^3\sigma {u_{d0}}\right)+2{k_r}^3\left(\delta+q^2\right)\left(u_{d0}+u_{i0}\right)\\[8pt]
\displaystyle  T_1=P_1\left({k_r}^4 {u_{i0}}^2 {u_{d0}}^2+3\sigma {k_r}^4 {u_{d0}}^2\right)-{k_r}^4 \left(\delta+q^2\right)\left({u_{i0}}^2 {u_{d0}}^2+3\sigma\right)
 \end{array}\right\}
\end{equation}
The imaginary dispersion relation is given by,
\begin{equation}\label{eq8}
    P_2 {\omega''}^4+Q_2 {\omega''}^3+R_2{\omega''}^2+S_2 {\omega''}+T_2=0,
\end{equation}
where, 
\begin{equation}\label{eq9}
\left. \begin{array}{ll}
\displaystyle P_2=-\left[{k_i}^2+\frac{1-2\kappa}{2\kappa-3}\right]
  \\[8pt]
  \displaystyle Q_2=P_2 {k_i}^2 \\[8pt]
  \displaystyle R_2=-P_2\left(3\sigma{k_i}^2-\eta_{ii}\eta_{di}{k_i}^4\right)-2{k_i}^2\left(\delta+q^2\right) \\[8pt]
  \displaystyle S_2=-3P_2 \sigma\eta_{di}{k_i}^4+{k_i}^4 \left(\delta+q^2\right)\left(\eta_{di}-\eta_{ii}\right) \\[8pt]
\displaystyle  T_2=-3\sigma {k_i}^4 \left(\delta+q^2\right)
 \end{array}\right\}
\end{equation}
\subsection{General Solution:}\label{sec3A}
These two equation (\ref{eq6}) and (\ref{eq8}) have the same form the solution of these kind of equation is given below;
The dispersion relations have the form
\begin{equation}\label{eq10}
    P \omega^4+Q \omega^3+R\omega^2+S \omega+T=0,
\end{equation}
or,
\begin{equation}\label{eq11}
     \omega^4+\frac{Q}{P} \omega^3+\frac{R}{P}\omega^2+\frac{S}{P}\omega+\frac{T}{P}=0,
\end{equation}
or,\begin{equation}\label{eq12}
     \omega^4+Q_3 \omega^3+R_3\omega^2+S_3\omega+T_3=0,
\end{equation}
Let, $\omega=x-\frac{Q_3}{4}$ then the above equation becomes,
\begin{equation}\label{eq13}
     x^4+p x^2+q x+r=0,
\end{equation}
where,
\begin{equation}\label{eq14}
\left. \begin{array}{ll}
\displaystyle p=-\frac{6{Q_3}^2}{16}-\frac{3Q_3}{4}+R_3
  \\[8pt]
  \displaystyle q=\frac{{Q_3}^3}{8}-\frac{2 Q_3 R_3}{4} +S_3 \\[8pt]
  \displaystyle r=-\frac{-3 {Q_3}^4}{256}+\frac{{Q_3}^2 R_3}{16}-\frac{Q_3}{4}+T_3 
 \end{array}\right\}
\end{equation}
From equation (\ref{eq13}) 
\begin{equation}\label{eq15}
    \left(x^2 +\frac{p}{2}\right)^2= -q x-r+\left(\frac{p}{2}\right)^2
\end{equation}
or,
\begin{equation}\label{eq16}
    \left(x^2 +\frac{p}{2}+u\right)^2= -q x-r+\left(\frac{p}{2}\right)^2+2 u x^2+p u+u^2
\end{equation}
In order to have perfect square, we have,
\begin{equation}\label{eq17}
    8 u^3+8 p u^2+ \left(2p^2-8r\right)u-q^2=0
\end{equation}
or,
\begin{equation}\label{eq18}
     u^3+ p u^2+ \left(\frac{2p^2-8r}{8}\right)u-\frac{q^2}{8}=0
\end{equation}
or, \begin{equation}\label{eq19}
    u^3 +p u^2 +l u+g=0,
\end{equation}
where, $l=\left(\frac{2p^2-8r}{8}\right)$ and $g=-\frac{q^2}{8}$.
\par
Again, let $u=t-\frac{p}{3}$ equation yields 
\begin{equation}\label{eq120}
    t^3 +j t+l=0,
\end{equation}
where, $j=\frac{3l-p^2}{3}$ and $l=\frac{2p^3-9pl+27g}{27}$.
Solution of this equation is,
\begin{equation}\label{eq21}
    t=\left(-\frac{l}{2}+\sqrt{\frac{l^2}{4}+\frac{j^3}{27}}\right)^{1/3}+\left(-\frac{l}{2}-\sqrt{\frac{l^2}{4}+\frac{j^3}{27}}\right)^{1/3}
\end{equation}
Now, we know $u=t-\frac{p}{3}$ the equation transforms into
\begin{equation}\label{eq22}
    x^2+\frac{p}{2}+u = \pm \left[\sqrt{2u}-\frac{2}{2\sqrt{2u}}\right]
\end{equation}
Solving this equation we get,
\begin{equation}\label{eq23}
    x=\pm \sqrt{-\frac{p}{2}-u \pm \left[\sqrt{2u}-\frac{2}{2\sqrt{2u}}\right]}
\end{equation}
Therefore we get,
\begin{equation}\label{eq24}
    \omega=-\frac{Q_3}{4}+x
\end{equation}
This equation (\ref{eq24}) is the general solution for the equations (\ref{eq6}) and (\ref{eq8}). Clearly this general solution has four roots that means the real dispersion relation (equation \ref{eq6}) and the imaginary dispersion relation (equation \ref{eq8}) individually have four roots. Under these four roots in each cases two roots are physically admissible \cite{Chaudhuriplasma}. Again in these two roots for each of the cases we can categorize them as `fast' and `slow' mode. 
\section{Derivation of KdV-Burgers Equation:}\label{sec4}
In order to derive the equation of motion for the nonlinear dust ion acoustic wave, we employ the reductive perturbation technique and define the following stretched variables,
\begin{equation}\label{eq25}
\left. \begin{array}{ll}
\displaystyle \xi = \varepsilon^{1/2}\left(x-Mt\right)
  \\[8pt]
  \displaystyle \tau = \varepsilon^{3/2}t \\[8pt]
  \displaystyle \eta = \varepsilon^{1/2}\eta_0
 \end{array}\right\}
\end{equation}
where $\varepsilon$ is a small parameter which characterizes the strength of nonlinearity, and $M$ is the Mach number i.e. the phase velocity of the wave. The stretching in $\eta$ is due to the small variations in perpendicular directions. 
\par
Now, Eqs (\ref{eq1}-\ref{eq4}) are written in terms of the stretched coordinates $\xi$, $\tau$, and $\eta$ and substituting the perturbation expansion given in Eq. (\ref{eq25}). From these equations equating the lowest orders in $\varepsilon$ (i.e., $\varepsilon^{3/2}$) with the boundary conditions that all variables, that is, ${n_i}^{(1)}$ , ${u_i}^{(1)}$, ${n_d}^{(1)}$ , ${u_d}^{(1)}$ and $\phi^{(1)}$ tend to zero as $\xi \xrightarrow[]{} \pm \infty$, we get,
\begin{equation}\label{eq26}
\left. \begin{array}{ll}
\displaystyle {u_d}^{(1)} = -\frac{q}{\left(M-u_{d0}\right)}\phi^{(1)}
  \\[8pt]
  \displaystyle {n_d}^{(1)} = -\frac{q}{\left(M-u_{d0}\right)^2}\phi^{(1)} \\[8pt]
  \displaystyle {u_i}^{(1)} = \frac{\phi^{(1)}\left(M-u_{di}\right)}{\left(M-u_{di}\right)^2-3\sigma}
  \\[8pt]
  \displaystyle {n_d}^{(1)} = \frac{\phi^{(1)}}{\left(M-u_{di}\right)^2-3\sigma}
 \end{array}\right\}
\end{equation}
Going to next higher order terms in $\varepsilon$, that is, $\varepsilon^{5/2}$-th term, we get,
\begin{equation}\label{eq27}
    \left(M-u_{d0}\right)\frac{\partial {n_d}^{(2)}}{\partial \xi}=\frac{\partial {u_d}^{(2)}}{\partial \xi}-\frac{q}{\left(M-u_{d0}\right)^2}\frac{\partial \phi^{(1)}}{\partial \tau}+\frac{q^2}{\left(M-u_{d0}\right)^3}\frac{\partial {\phi^{(1)}}^2}{\partial \xi}
\end{equation}
\begin{dmath}\label{eq28}
    \left(M-u_{i0}\right)\frac{\partial {n_i}^{(2)}}{\partial \xi}=\frac{\partial {u_i}^{(2)}}{\partial \xi}-\frac{1}{\left[\left(M-u_{i0}\right)^2-3\sigma\right]}\frac{\partial \phi^{(1)}}{\partial \tau}+\frac{\left(M-u_{i0}\right)}{\left[\left(M-u_{i0}\right)^2-3\sigma\right]}\frac{\partial {\phi^{(1)}}^2}{\partial \xi}
 \end{dmath}
 
\begin{dmath}\label{eq29}
    \frac{\partial {u_d}^{(2)}}{\partial \xi}=-\frac{q}{\left(M-u_{d0}\right)^2}\frac{\partial \phi^{(1)}}{\partial \tau}+\frac{q^2}{\left(M-u_{d0}\right)^3}\frac{\partial \phi^{(1)}}{\partial \xi}-\frac{q}{\left(M-u_{d0}\right)^2}\frac{\partial \phi^{(2)}}{\partial \xi}+\frac{q\eta_{d0}}{\left(M-u_{d0}\right)^2}\frac{\partial^2 \phi^{(1)}}{\partial \xi^2}
\end{dmath}
\begin{dmath}\label{eq30}
    \frac{\partial {u_i}^{(2)}}{\partial \xi}=\frac{1}{M-u_{i0}}\frac{\partial \phi^{(2)}}{\partial \xi}+\frac{1}{\left[\left(M-u_{i0}\right)^2-3\sigma\right]}\frac{\partial \phi^{(1)}}{\partial \tau}+\frac{\left[ 3\left(M-u_{i0}\right)^2-3\sigma\right]}{\left(M-u_{i0}\right)\left[\left(M-u_{i0}\right)^2-3\sigma\right]^2}\phi^{(1)}\frac{\partial \phi^{(1)}}{\partial \xi}+\frac{3 \sigma}{\left(M-u_{i0}\right)}\frac{\partial {n_i}^{(2)}}{\partial \xi}+\frac{\eta_{i0}}{\left[\left(M-u_{i0}\right)^2-3\sigma\right]}\frac{\partial^2 \phi^{(1)}}{\partial \xi^2}
\end{dmath}
The second-order perturbation of the Poisson’s equation can be equated as
\begin{dmath}\label{eq31}
    \frac{\partial^2 \phi^{(1)}}{\partial \xi^2}=\left[\frac{1-2\kappa}{2\kappa-3}+\frac{\left(4\kappa^2-1\right)}{\left(2\kappa-3\right)^2}\right]\phi^{(2)}+\frac{\left(4\kappa^2-1\right)}{2\left(2\kappa-3\right)^2}{\phi^{(1)}}^2-\delta {n_i}^{(2)}+q{n_d}^{(2)}
\end{dmath}
Differentiating both sides of Eq. (\ref{eq31}) by $\xi$ and carrying out a detailed algebraic treatment with the Equations from (\ref{eq27}) to (\ref{eq30}), the nonlinear KdV–Burgers equation is given by the following equation
\begin{equation}\label{eq32}
    \frac{\partial \phi^{(1)}}{\partial \tau}+A\phi^{(1)}\frac{\partial\phi^{(1)}}{\partial \xi}+B\frac{\partial^3\phi^{(1)}}{\partial \xi^3}+C\frac{\partial^2\phi^{(1)}}{\partial \xi^2}=0
\end{equation}
where, $A=\frac{\Gamma}{\Theta}$, $B=\frac{1}{\Theta}$ and $C=\frac{\Upsilon}{\Theta}$. The $\Gamma$, $\Theta$ and $\Upsilon$ are given below.
\begin{equation}\label{eq33}
\left. \begin{array}{ll}
\displaystyle \Gamma=\frac{3\delta\left[\left(M-u_{i0}\right)+\sigma\right]}{\left[\left(M-u_{i0}\right)^2-3\sigma\right]^3}-\frac{3q^3}{\left(M-u_{d0}\right)^4}-\frac{4\kappa^2-1}{\left(2\kappa-3\right)^2}
  \\[8pt]
  \displaystyle \Theta =\frac{2\left(M-u_{i0}\right)\delta}{\left[\left(M-u_{i0}\right)^2-3\sigma\right]^2}+\frac{2q^2}{\left(M-u_{d0}\right)^2} \\[8pt]
  \displaystyle \Upsilon=\frac{\eta_{i0}\left(M-u_{i0}\right)\delta}{\left[\left(M-u_{i0}\right)^2-3\sigma\right]^2}-\frac{\eta_{d0}q^2}{\left(M-u_{d0}\right)^3}
 \end{array}\right\}
\end{equation}
The nonlinear equation (\ref{eq32}) is the well known KdV-Burgers(KDVB) equation and the coefficients $A$, $B$ and $C$ are the nonlinear, dispersive and dissipative coefficients, respectively. 
\subsection{Solution of KdVBurgers Equation: }\label{sec4A}
The equation (\ref{eq32}) is nonlinear equation in both the variables $\xi$ a space like variable and $\tau$ a time like variable. To solve this equation we have to first join these variables into one wave variable $\psi=\xi-M\tau$ to transform the partial differential equation (PDE) of two variables $\xi$ and $\tau$ into ordinary differential equation (ODE) of one variable $\psi$ with the application of the boundary conditions when $\psi \xrightarrow[]{}0$ then $\phi^{(1)} \xrightarrow[]{} 0$ and $\frac{\partial \phi^{(1)}}{\partial \psi} \xrightarrow[]{} 0$ as \cite{goswami2019shock, goswami2020collision, goswami2020electron}
\begin{equation}\label{eq34}
    -M\frac{d\phi^{(1)}}{d\psi}+A\phi^{(1)}\frac{d\phi^{(1)}}{d\psi}+B\frac{d^2\phi^{(1)}}{d\psi^2}+C\frac{d^3\phi^{(1)}}{d\psi^3}
\end{equation}
As we are trying to solve this equation with the help of tan-hyperbolic method so now we have to take $s=tanh\psi$ and assuming a series solution predicted by Wazwaz \cite{WAZWAZ2008584} as $\phi^{(1)}(s)=\sum_{j=0}^{n}a_j s^j$ and we get the following solution,
\begin{equation}\label{eq35}
    \phi^{(1)}=\frac{12A}{B}\left[1-tanh^2(\psi)\right]-\frac{36C}{15A}tanh(\psi)
\end{equation}
This equation (\ref{eq35}) is the solution of the KdVB equation (\ref{eq32}).
\subsection{The KdV equation and the solution:}\label{sec4B}
It is very clear that if there is no viscous force considered for both the species, i.e. ion and dust then $\eta_{i0}=\eta_{d0}=0$. So there will be no dissipation and $C=0$. In this condition, the equation (\ref{eq32}) is transformed into Kortewg deVries (KdV) equation.
\begin{equation}\label{eq36}
    \frac{\partial \phi^{(1)}}{\partial \tau}+A\phi^{(1)}\frac{\partial\phi^{(1)}}{\partial \xi}+B\frac{\partial^3\phi^{(1)}}{\partial \xi^3}
\end{equation}
The coefficients $A$ and $B$ are already given. Using the same transformation $\psi=\xi-M\tau$ and the boundary condition we have the solution of the KdV equation,
\begin{equation}\label{eq37}
    \phi^{(1)}=\phi_0 sech^2\left(\frac{\psi}{\Delta}\right)
\end{equation}
where, $\phi_0=\frac{3M}{A}$ and $\Delta=2\sqrt{\frac{M}{C}}$
\subsection{Numerical solution of KdV:}\label{sec4C}
Following the same procedure as Soliman, Ali and Raslan \cite{SOLIMAN20091107}, we find out the numerical solution of the KdV equation. Taking $F=A\frac{{\phi^{(1)}}^{2}}{2}+B\frac{\partial^2 \phi^{(1)}}{\partial \xi^2}$ the equation (\ref{eq36}) transforms into the given form
\begin{equation}\label{eq38}
    \frac{\partial \phi^{(1)}}{\partial \tau}+\frac{\partial F}{\partial \xi}=0
\end{equation}
Using the finite difference method the equation (\ref{eq38}) becomes,
\begin{equation}\label{eq39}
    \frac{{\phi_i{^{(1)}}^{n+1}-\phi_i{^{(1)}}}^{n}}{\Delta \tau}+\frac{F_{i+\frac{1}{2}}-F_{i-\frac{1}{2}}}{\Delta \xi}=0
\end{equation}
Using the similarity solution procedure we can have the solution of equation (\ref{eq36}) as follows (representing the $\phi^{(1)}$ as $\phi$ here),
\begin{dmath}\label{eq40}
    \phi_i{^{{n+1}}}=\phi_i{^{n}}-\frac{\Delta \tau}{\Delta \xi}\left[\frac{A}{4}\left\{\left(\phi_{i+1}-(\Phi_x)_{i+1}H+(\Phi_{xx})_{i+1}M_1\right)^2+\left(\phi_{i}+(\Phi_x)_{i}H+(\Phi_{xx})_{i}M_1\right)^2-\left(\phi_{i}-(\Phi_x)_{i}H+(\Phi_{xx})_{i}M_1\right)^2-\left(\phi_{i-1}-(\Phi_x)_{i-1}H+(\Phi_{xx})_{i-1}M_1\right)^2\right\}+\frac{B}{2}\left\{b^2(\Phi_x)_{i+1}H+(\Phi_{xx})_{i+1}H+(\Phi_{xx})_{i+1}M_2-b^2(U_x)_{i}H+(\Phi_{xx})_{i}M_2-b^2(\Phi_x)_i H-(\Phi_{xx})_{i+1}M_2+b^2(\Phi_x)_{i-1}H-(\Phi_{xx})_{i-1} M_2 \right\}\right]
\end{dmath}
Here,
\begin{equation}\label{eq41}
   \left. \begin{array}{ll}
\displaystyle H=\frac{1}{4cos\alpha}
  \\[8pt]
  \displaystyle M_1=-\frac{cos\alpha-1}{4sin\alpha} \\[8pt]
  \displaystyle M_2=\frac{b^2cos\alpha}{4sin^2\alpha}
 \end{array}\right\}
\end{equation}
where,
\begin{equation}\label{eq42}
   \left. \begin{array}{ll}
\displaystyle \alpha=\frac{1}{2}b\Delta \xi
  \\[8pt]
  \displaystyle \beta=b M \Delta \tau 
 \end{array}\right\}
\end{equation}
also \begin{equation}\label{eq43}
    b=\sqrt{\frac{M+AD}{B}}
\end{equation}
and
also \begin{equation}\label{eq44}
    \left. \begin{array}{ll}
  \displaystyle \Phi_{xx}={\phi^{n}}_{i+1}-2{\phi^{n}}_{i}+{\phi^{n}}_{i-1} \\[8pt]
  \displaystyle \Phi_x={\phi^{n}}_{i+1}-{\phi^{n}}_{i-1}
 \end{array}\right\}
\end{equation}
Here, $D$ is the local similarity equation.
\section{Phase Plane Analysis:}\label{sec5}
In this section, we transform the KdV equation (\ref{eq36}) into a dynamical system with the help of $\chi=\xi-v\tau$ and $\phi^{(1)}(\xi,\tau)=\phi^{(1)}(\chi)$
\begin{equation}\label{eq45}
    \left. \begin{array}{ll}
  \displaystyle \frac{d \phi^{(1)}}{d\chi}=z \\[8pt]
  \displaystyle \frac{dz}{d\chi}=\frac{v}{B}\phi^{(1)}-\frac{A}{B}{\phi^{(1)}}^2
 \end{array}\right\} .
\end{equation}
\par
To locate critical points by solving the equations $\Dot{\phi}=\Dot{z}=0$, ` $\Dot{}$ ' sign represents $\frac{d}{d\chi}$.
\par
Hence, $\Dot{\phi}=0$ if $z=0$ and $\Dot{z}=0$ if $\left(\frac{v}{B}\phi -\frac{A}{B}\phi^2\right)=0$.
\par
If $z=0$ then $\Dot{z}=0$ if $\left(\frac{v}{B}\phi -\frac{A}{B}\phi^2 \right)=0$ which has two solutions, $\phi=0$ and $\phi=\frac{v}{A}$. Therefore, there are two critical points $(0,0)$ and $(\frac{v}{A},0)$.
\par
Linearize by finding the Jacobian matrix; hence,
\begin{equation}\label{eq46}
    J=\begin{pmatrix}
0 & 1\\
\frac{v}{B}-\frac{2A}{B}\phi & -\frac{C}{B}
\end{pmatrix}
\end{equation}
Linearizing in first critical point that is $(0,0)$ we get,
\begin{equation}\label{eq47}
    J_{(0,0)}=\begin{pmatrix}
0 & 1\\
\frac{v}{B} & -\frac{C}{B}
\end{pmatrix}
\end{equation}
The Eigen values for this Jacobian are $\lambda = \frac{-\frac{C}{B} \pm \sqrt{\frac{C^2}{B^2}+\frac{4v}{B}}}{2}$.
Again, linearizing in second critical point that is $(\frac{v}{A},0)$ we get,
\begin{equation}\label{eq48}
    J_{(\frac{v}{A},0)}=\begin{pmatrix}
0 & 1\\
-\frac{v}{B} & -\frac{C}{B}
\end{pmatrix}
\end{equation}
And the eigenvalues for this Jacobian are $\lambda = \frac{-\frac{C}{B} \pm \sqrt{\frac{C^2}{B^2}-\frac{4v}{B}}}{2}$.
\par
From the eigenvalues at the two critical points it is clear that for the critical point $(0,0)$ the eigenvalues are always real as the dispersive coefficient $B$ is always positive in this model but there is a possibility that the eigenvalues for the critical point $(\frac{v}{A},0)$ can be imaginary if $\frac{4v}{B}>\frac{C^2}{B^2}$. 
\par
The dynamical system \ref{eq45} are Hamiltonian system are Hamiltonian system of the Hamiltonian function,
\begin{equation}\label{eq49}
    H(\phi, z)=\frac{z^2}{2}-\frac{v \phi^2}{2B}-\frac{A \phi^3}{3B}
\end{equation}
This is obviously a two dimensional Hamiltonian function.
\par
From the nature of the Jacobian and the eigenvalues, we categorize these critical points in two distinct category.
\par
(i) At equilibrium point or critical point $E_0$
\begin{equation}\label{eq50}
D = det J_{(0,0)}=
\begin{vmatrix}
0 & 1  \\ 
\frac{v}{B} & -\frac{C}{B} 
\end{vmatrix}
=
-\frac{v}{B}<0
\end{equation}
which implies $E_0$ is a saddle point.
\par
(ii) At equilibrium/ critical point $E_1$ i.e. $(\frac{v}{A},0)$
\begin{equation}\label{eq51}
D = det J_{(0,0)}=
\begin{vmatrix}
0 & 1  \\ 
-\frac{v}{B} & -\frac{C}{B} 
\end{vmatrix}
=
\frac{v}{B}>0
\end{equation}
which implies $E_1$ is a center.
\par
The trace of both the determinate is same which is $-\frac{C}{B}>0$, as $C$ itself is a negative quantity.
\section{Results:}\label{sec6}
\begin{figure}
    \centering
     \includegraphics[width=10cm]{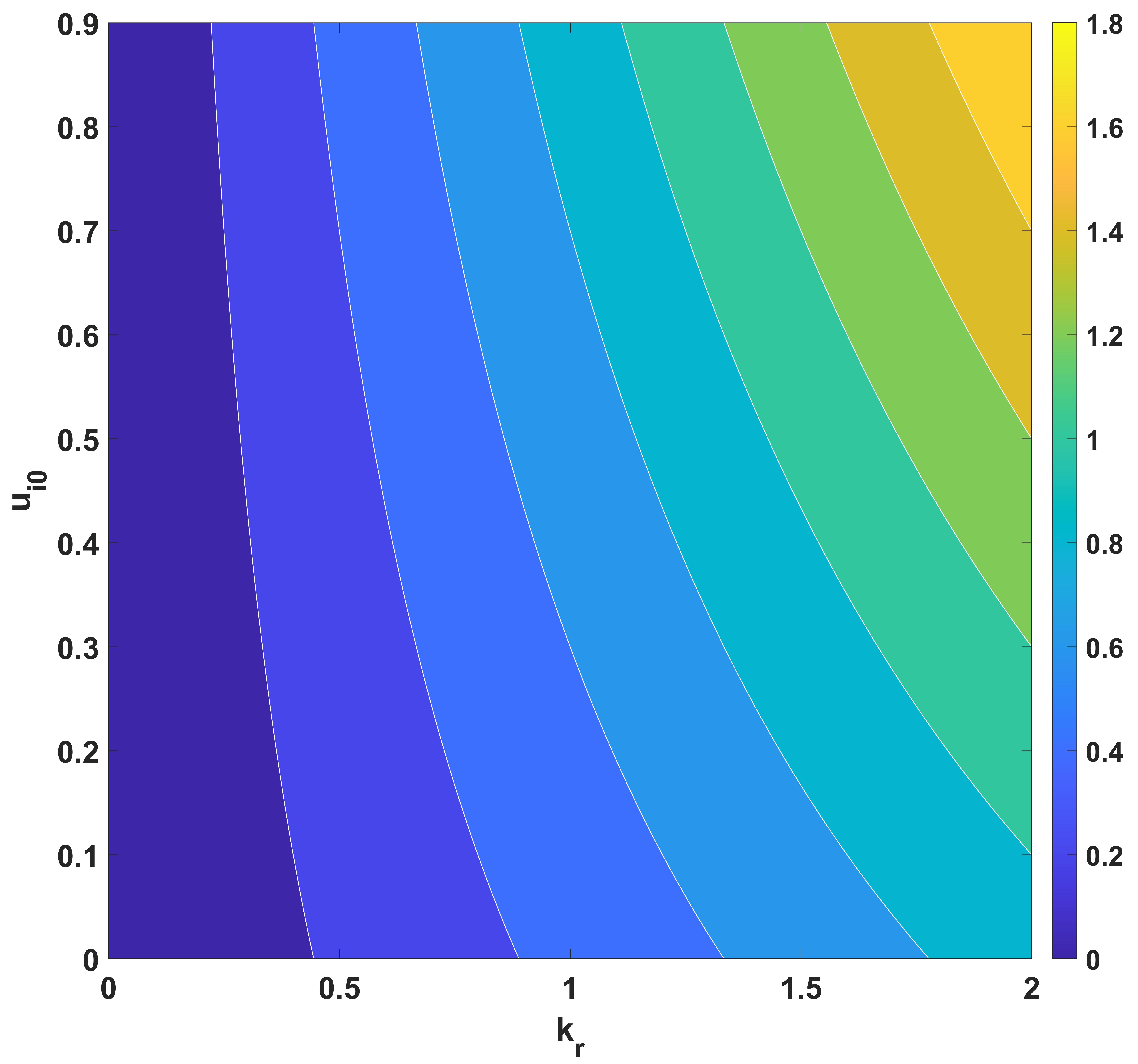}
    \caption{Contour plot of one root of $\omega'$, carrier wave number $k_r$ and unperturbed ion streaming velocity $u_{i0}$ with unperturbed dust streaming velocity $(u_{d0}) = 0.07$, finite temperature coefficient $(\sigma)=0.65$, kappa index $(\kappa)=5$ and dust charge $(q)=1$}
    \label{realdispersionrelation3}
\end{figure}
\begin{figure}
    \centering
     \includegraphics[width=10cm]{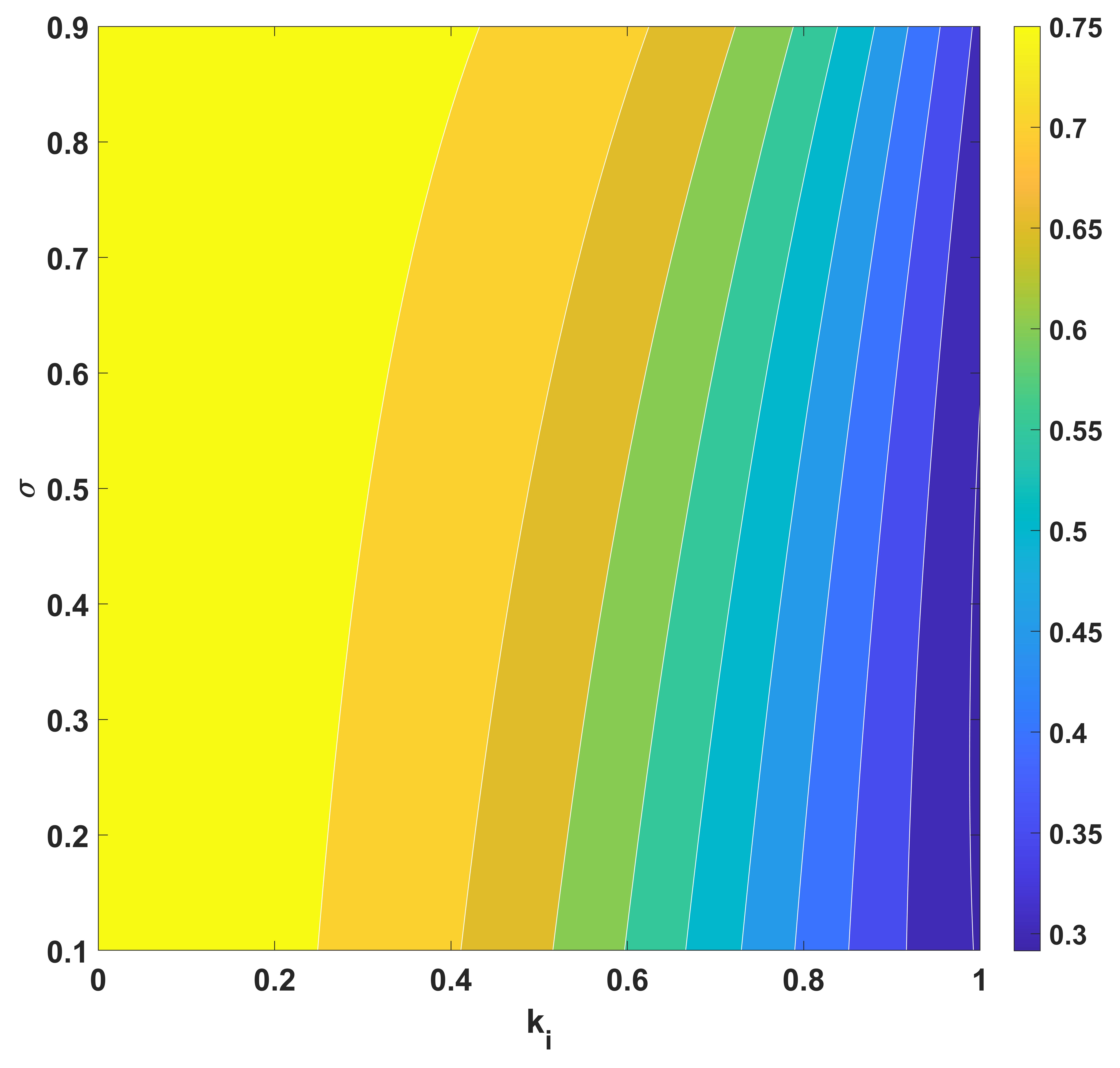}
    \caption{Contour plot of one root of $\omega''$, carrier wave number $k_i$ and finite temperature coefficient $(\sigma)$ with kinematic viscosity of ion $(\eta_{i0})=0.09$, kinematic viscosity of dust $(\eta_{d0})=0.07$, finite temperature coefficient $(\sigma)=0.65$, kappa index $(\kappa)=5$ and dust charge $(q)=1$}
    \label{imagainarydispersion3}
\end{figure}
\begin{figure}
    \centering
     \includegraphics[width=10cm]{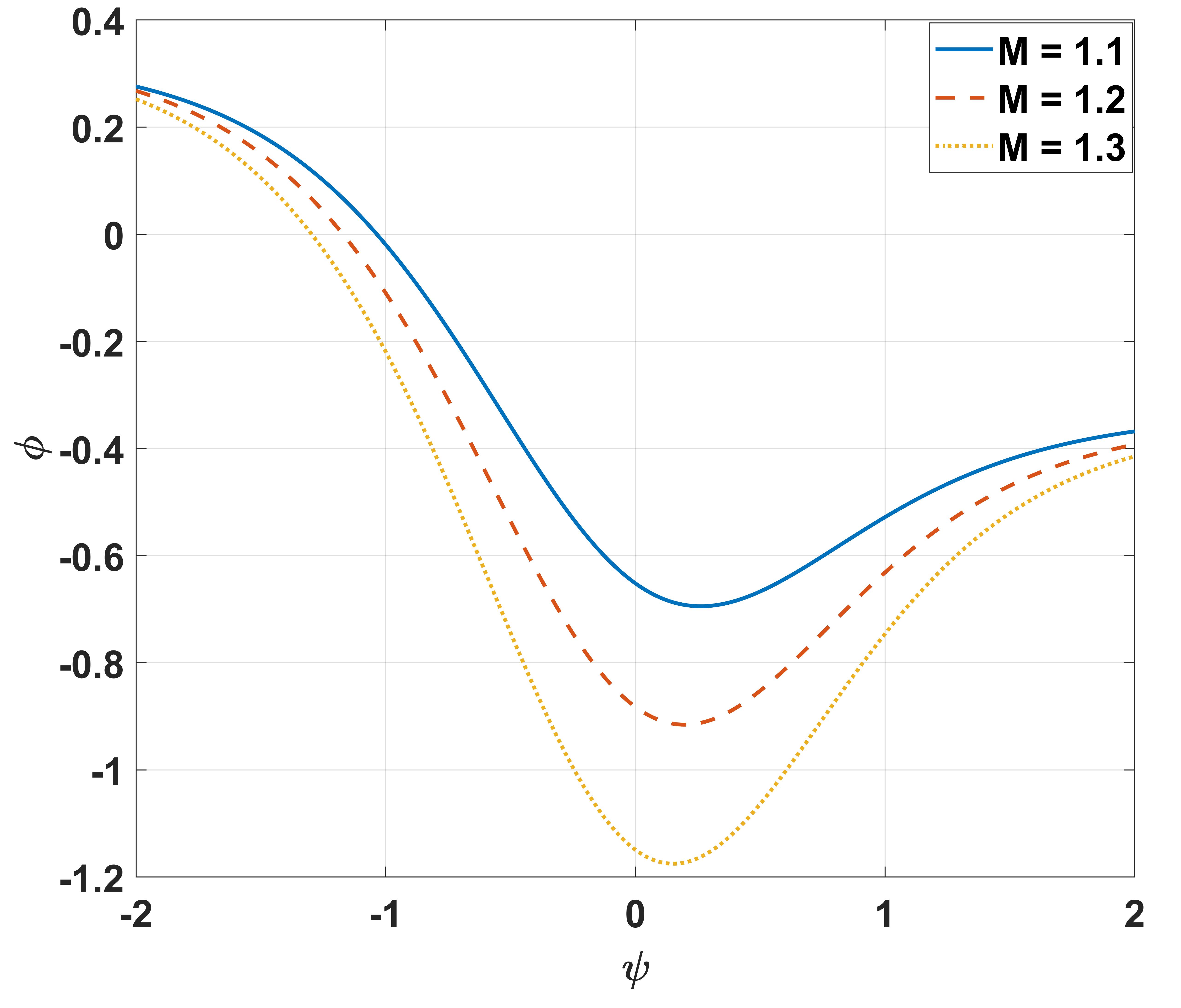}
    \caption{Shock for different Mach numbers $(M)$ with unperturbed ion streaming velocity$(u_{i0})=0.3$, unperturbed dust streaming velocity$(u_{d0})=0.07$, kinematic viscosity of ion $(\eta_{i0})=0.09$, kinematic viscosity of dust $(\eta_{d0})=0.07$, finite temperature coefficient $(\sigma)=0.65$, kappa index $(\kappa)=5$ and dust charge $(q)=1$}
    \label{shockfordifferentmachnumber}
\end{figure}

\begin{figure}
    \centering
    \includegraphics[width=10cm]{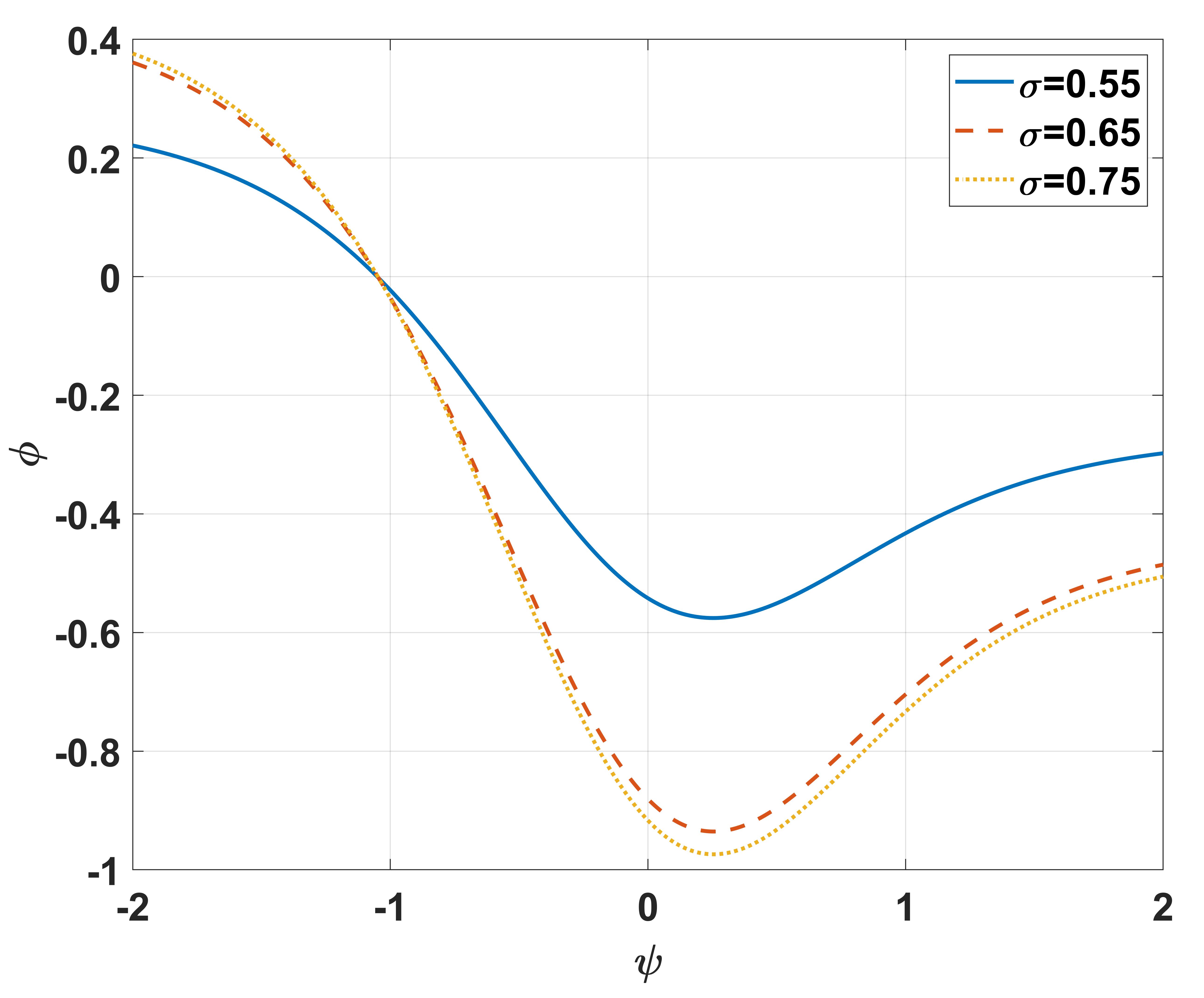}
    \caption{Shock for different finite temperature coefficients $(\sigma)$ with unperturbed ion streaming velocity$(u_{i0})=0.3$, unperturbed dust streaming velocity$(u_{d0})=0.07$, kinematic viscosity of ion $(\eta_{i0})=0.09$, kinematic viscosity of dust $(\eta_{d0})=0.07$, Mach number $(M)=1.2$, kappa index $(\kappa)=5$ and dust charge $(q)=1$}
    \label{shockfordifferentsigma}
\end{figure}
\begin{figure}
    \centering
     \includegraphics[width=10cm]{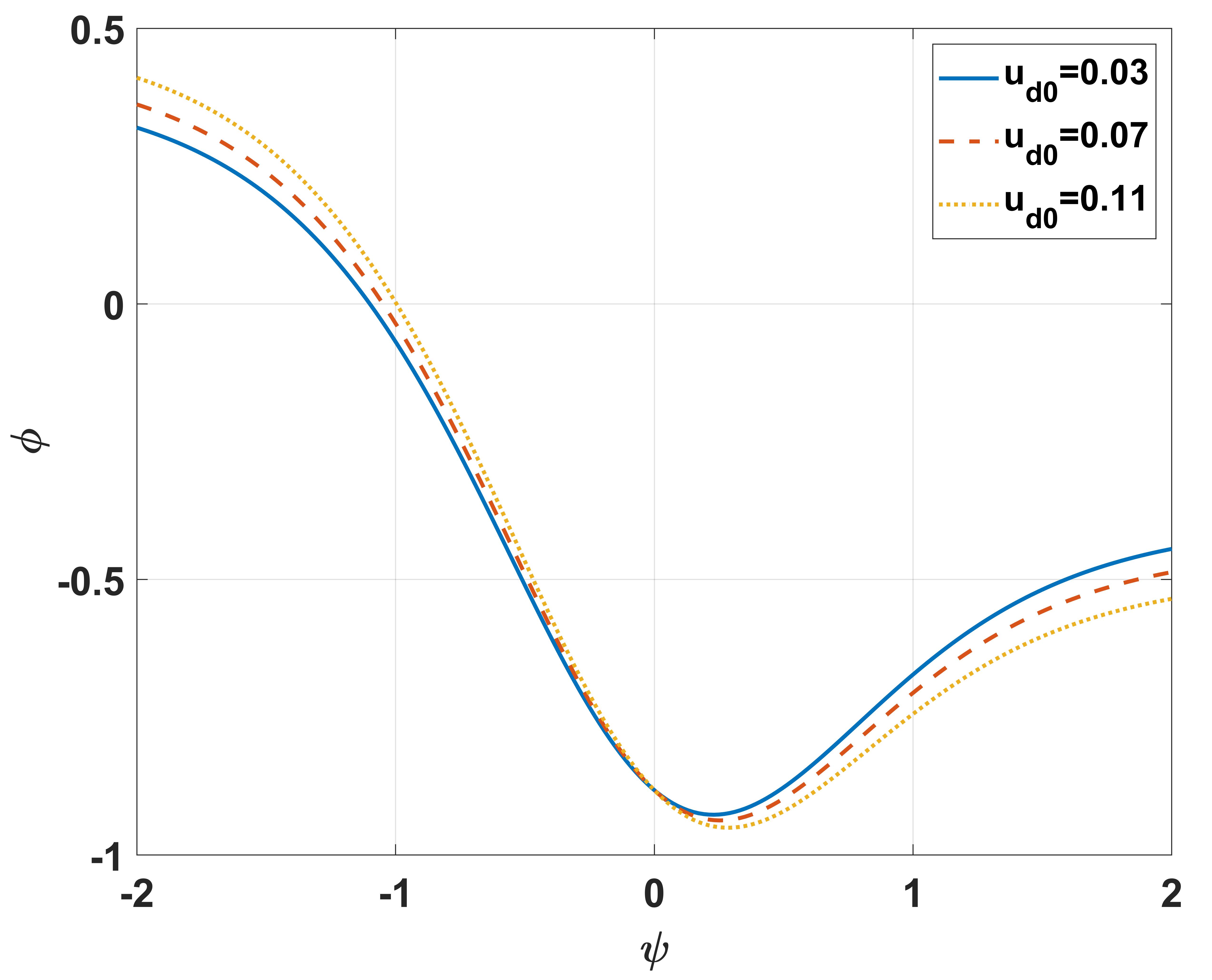}
    \caption{Shock for different unperturbed dust streaming velocities $(u_{d0})$  with unperturbed ion streaming velocity$(u_{i0})=0.3$, kinematic viscosity of dust $(\eta_{d0}) =0.07$, kinematic viscosity of ion $(\eta_{i0})=0.09$, finite temperature coefficient $(\sigma)=0.65$, Mach number $(M)=1.2$, kappa index $(\kappa)=5$ and dust charge $(q)=1$}
    \label{shockfordifferentstreamingofdust}
\end{figure}
\begin{figure}
    \centering
     \includegraphics[width=10cm]{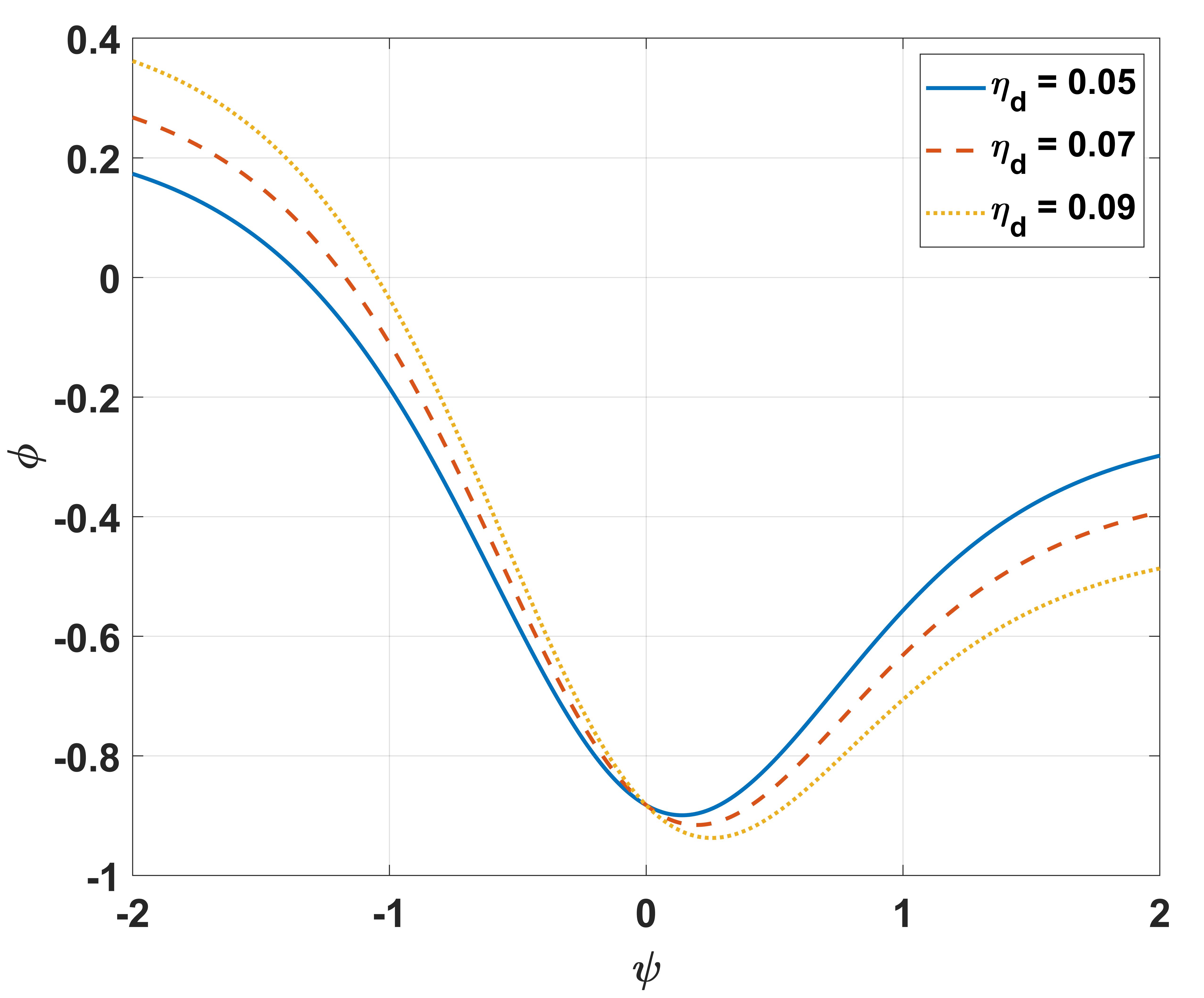}
    \caption{Shock for different kinematic viscosity of dust $(\eta_{d0})$  with unperturbed ion streaming velocity$(u_{i0})=0.3$, unperturbed dust streaming velocity$(u_{d0})=0.07$, kinematic viscosity of ion $(\eta_{i0})=0.09$, finite temperature coefficient $(\sigma)=0.65$, Mach number $(M)=1.2$, kappa index $(\kappa)=5$ and dust charge $(q)=1$}
    \label{shockfordifferentetad}
\end{figure}
\begin{figure}
    \centering
     \includegraphics[width=10cm]{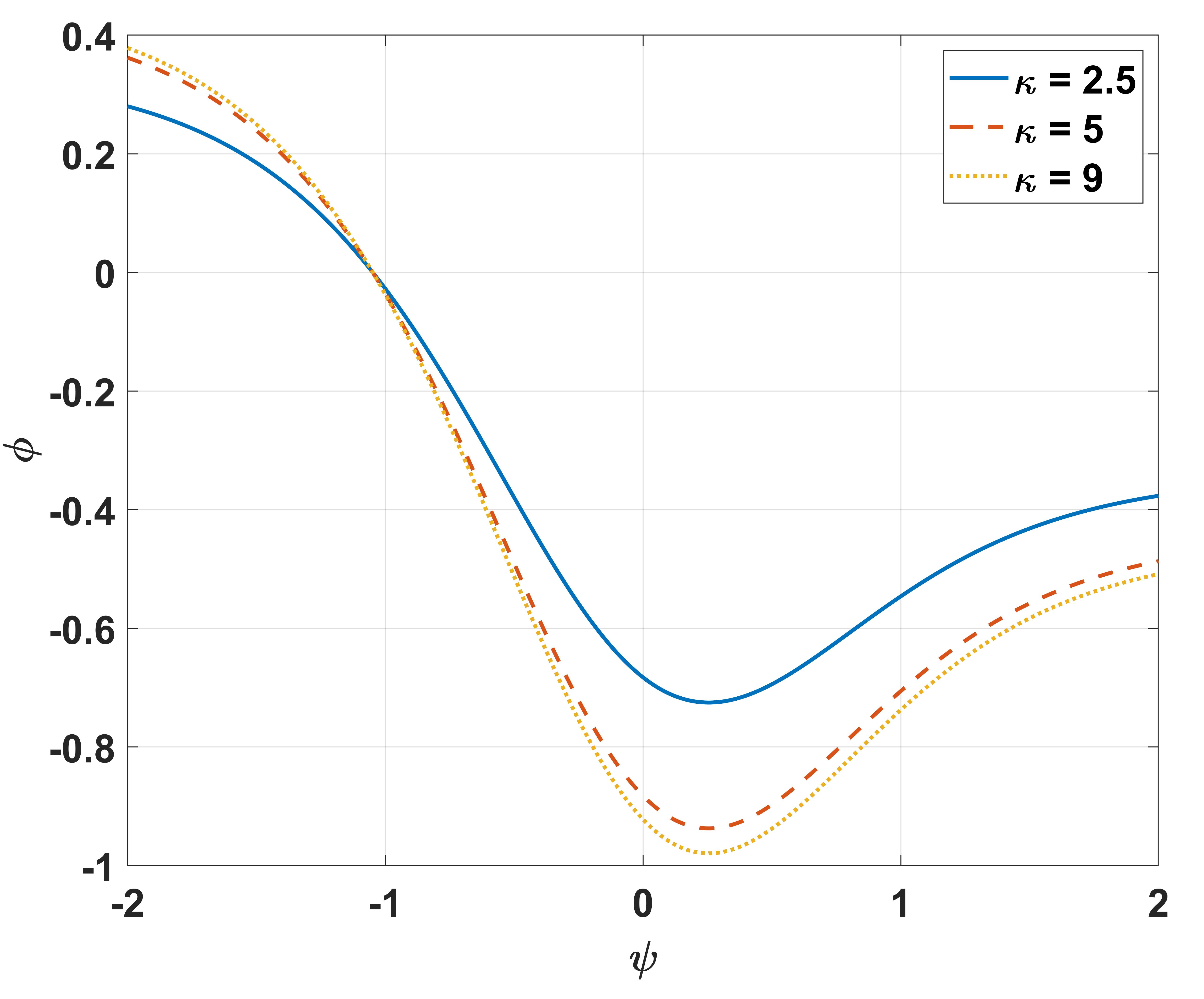}
    \caption{Shock for different kappa indices $(\kappa)$  with unperturbed ion streaming velocity$(u_{i0})=0.3$, unperturbed dust streaming velocity$(u_{d0})=0.07$, kinematic viscosity of dust $(\eta_{d0}) =0.07$, kinematic viscosity of ion $(\eta_{i0})=0.09$, finite temperature coefficient $(\sigma)=0.65$, Mach number $(M)=1.2$, and dust charge $(q)=1$}
    \label{shockfordifferentkappa}
\end{figure}
\begin{figure}
    \centering
     \includegraphics[width=10cm]{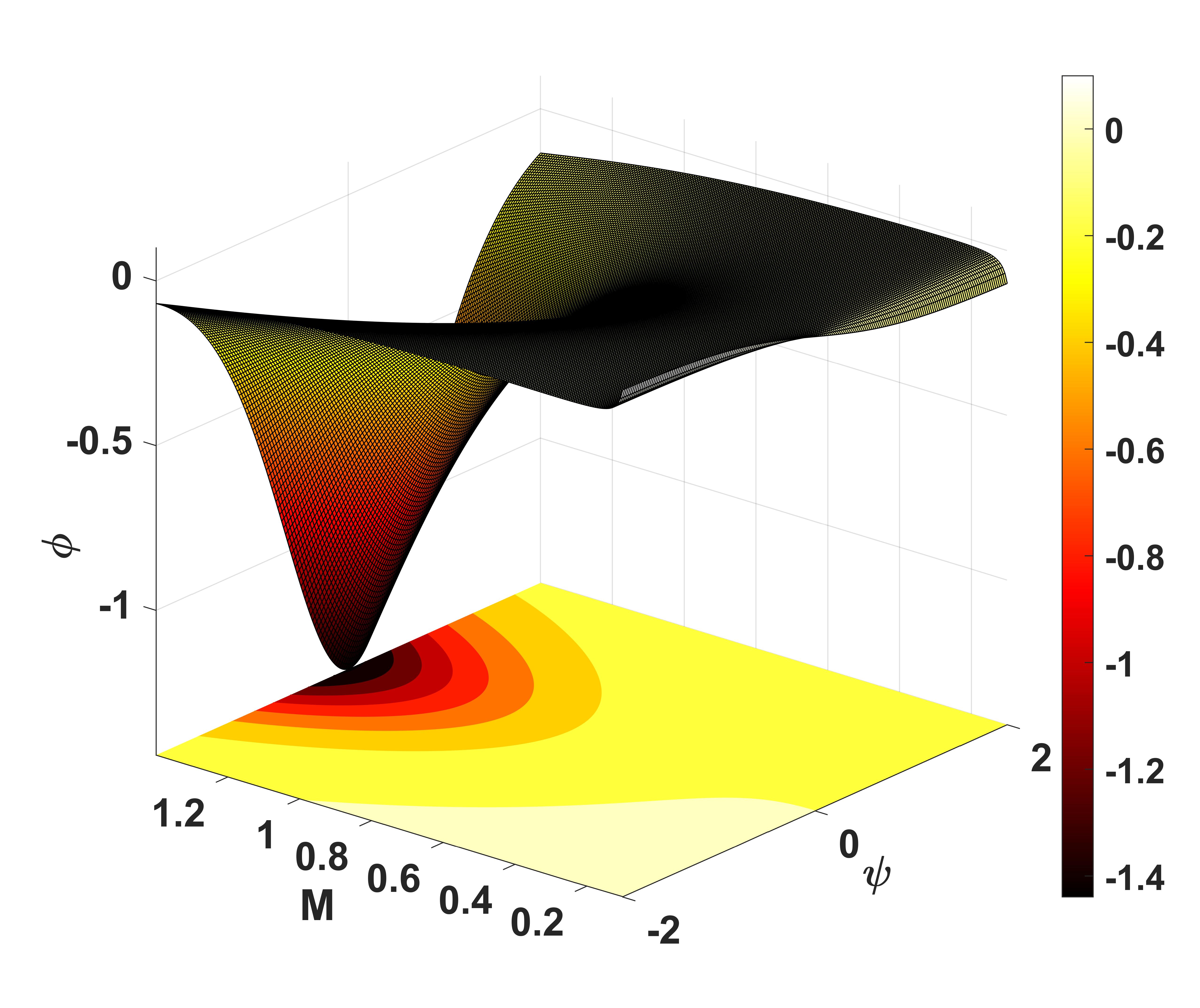}
    \caption{Transition from Weak to Strong Shock Region}
    \label{transitionfromweaktostrongshock}
\end{figure}
\begin{figure}
    \centering
     \includegraphics[width=10cm]{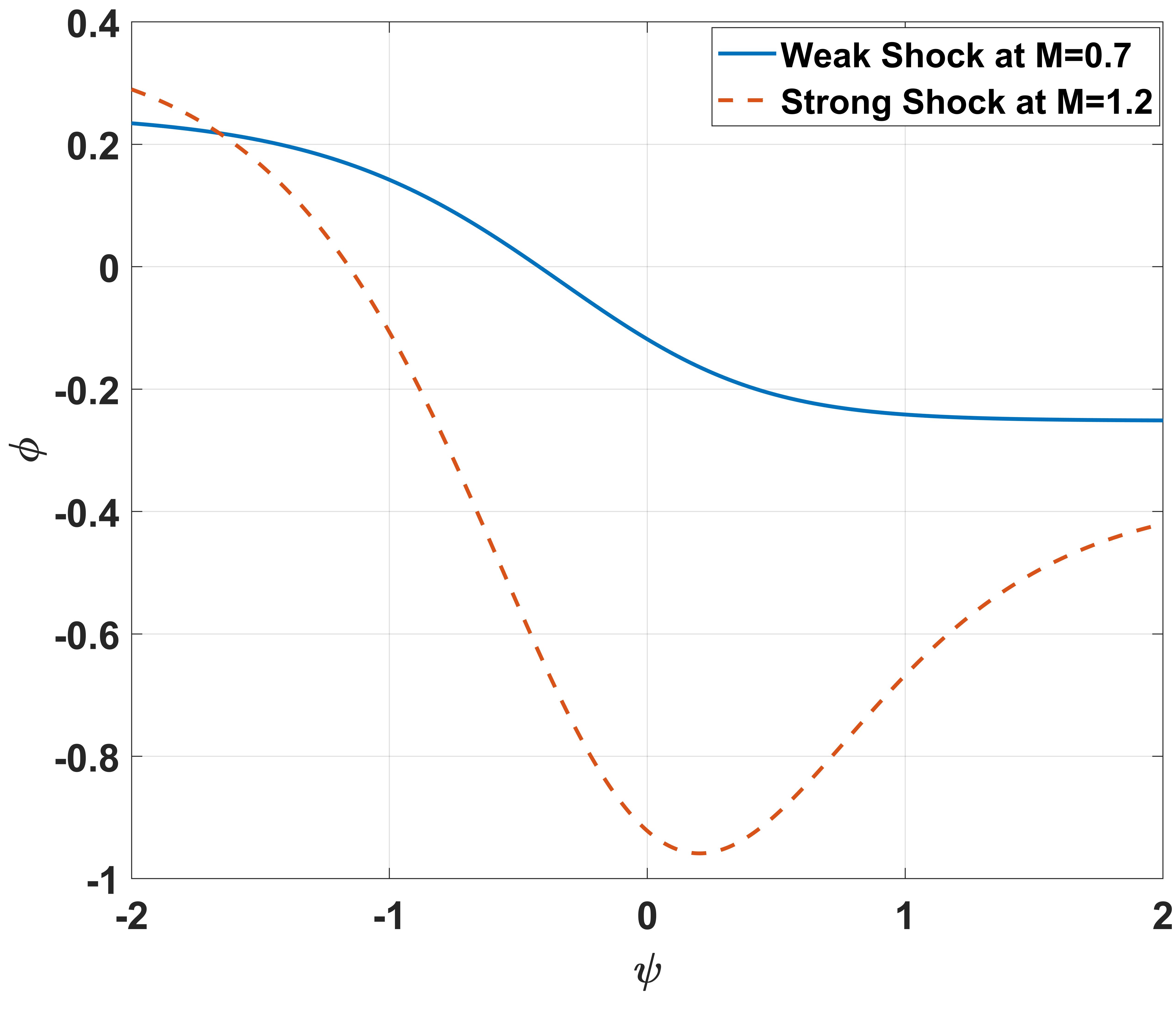}
    \caption{Weak and Strong Shock}
    \label{Weakandstrongshock}
\end{figure}
In the section \ref{sec4}, we have derived and analyse various linear and nonlinear effect in this plasma model analytically and numerically. At first we will discuss about the variation of the one root of the real dispersion relation with various ion streaming velocity in the figure \ref{realdispersionrelation3}.
In this contour plot we can see that the nonlinear damping rate increases as the unperturbed ion velocity $(u_{i0})$ increases. Next, in the figure \ref{imagainarydispersion3} we have discussed the variation of one imaginary root of the imaginary dispersion relation expressed in the equation \ref{eq8}. This is also a contour plot but the nature of it is totally different from the previous one (figure \ref{realdispersionrelation3}). This change in nature have two possible reasons. Firstly, in the calculation of imaginary dispersion relation \ref{eq8} we do not consider the streaming velocity of ions and the dust grains $u_{i0}=u_{d0}=0$. And, secondly we have considered the viscous coefficients for both the species. Basically, the kinematic viscosity is the origin of this type of dispersion relation and with the increase of the viscous coefficient the value of $\omega''$ will increase with the increase of $k_i$. Also from figure \ref{imagainarydispersion3} we can say that the finite temperature coefficient of ions $\sigma$ plays a pivotal role in the imaginary dispersion relation with the increase of the finite temperature coefficient of ions the value of $\omega''$ will decrease with the increase of $k_i$.  
\par
Now, in this section we will discuss parametric dependence of the different wave structures. Firstly, in figure \ref{shockfordifferentmachnumber}, we discuss the variation of the shock wave for different Mach number $(M)$ but all the other plasma quantities are constant. When the Mach number, i.e., the value of the frame velocity increases the width and the amplitude of the shock fronts increase. It is clear that the nonlinear and dissipative coefficients $A$ and $C$ have negative values, and the value of these decreases with the growth of the Mach number. On the other hand, the dispersive coefficient $B$ is positive, and this coefficient $B$ increases with the increment of the Mach number. This result can also explained physically as when the wave frame velocity is high that means the dust-ion acoustic waves propagate faster in the medium then it can have very little time to form the non-linearity but the medium also has the little time to apply the existing dissipation on it. Whereas the fastness of that same waves increases the disperive nature of the medium. In figure \ref{shockfordifferentsigma}, we discuss the variation of the shock wave for different finite temperature coefficient $(\sigma)$ with all the other quantities remain constant. This is basically a degeneracy parameter and discussion of it is little bit difficult. Here, the increase of this parameter $\sigma$ the non-linearity of the shock waves decreases but the dispersive and disipation coefficient both increases. But it is clearly seen that for lower value of the degeneracy parameter $\sigma$ has very significant effect in all these three parameters but with the increase of the value of $\sigma$ the dependence decreases very sharply. For an example, we can say that with the values of the other plasma parameters given in the figure \ref{shockfordifferentsigma}, the change of the parameter $\sigma$ from $0.55$ to $0.65$, the nonlinear, dispersive, and dissipation coefficients change from $-6.1131$ to $-2.6796$, $0.0539$ to $0.1504$, and $-0.0721$ to $-0.2818$ but when the parameter $\sigma$ from $0.65$ to $0.75$ the nonlinear, dispersive, and dissipation coefficients change from $-2.6796$ to $-2.6496$, $0.1504$ to $0.1802$, and $-0.2818$ to $-0.3464$. This happens as we consider the supersonic shock region (discussed in the discussion of figure \ref{Weakandstrongshock}) and the unperturbed ion velocity is higher than the unperturbed dust velocity so the effect of degeneracy parameter $\sigma$  become less significant when the value of it is sufficiently higher than the critical value i.e. $\sigma_c= \frac{\left(M-u_{i0}\right)^2}{3}$. Next we will discuss about the change of the shock profile for various unperturbed velocities of the charge negative dust particles and it is given in the figure \ref{shockfordifferentstreamingofdust}. Dust particles are comparatively heavier than that of the ions. In this whole discussion we have considered the ion velocity more or less than ten times greater than the velocity of dust particles. As we said in the previous result that we have discussed most of the results in the supersonic region so the effect of the dust streaming velocity is very less but it is the opposite of the result that previously found for electro-static\cite{goswami2020collision} or electron-acoustic \cite{goswami2019shock,goswami2020electron} cases. In these previous cases, when the streaming velocity of dust increases the nonlinearity decreases and dispersive effect increases significantly but in this case though the nonlinearity decreases with the increase of streaming velocity of dust particles but this change is not significant and the effect of this is contoured by decrease of the dispersive coefficient and also the increase of dissipative coefficient in formation of the shock profile. Now we will discuss about the change in the shock profile due to change in the kinematic viscosity of dust particle in figure \ref{shockfordifferentetad}. It is very clear from eq. \ref{eq33} that there will be no change in $A$ or $B$ due to change in viscous coefficient the only changes happen in the coefficient $C$. The dust viscous coefficient is the main reason of the shock formation in a dust-ion acoustic waves because in light of physically acceptable values in supersonic wave region $\frac{\eta_{i0}\left(M-u_{i0}\right)\delta}{\left[\left(M-u_{i0}\right)^2-3\sigma \right]^2}<\frac{\eta_{d0}q^2}{\left(M-u_{d0}\right)^3}$. In figure  \ref{shockfordifferentkappa} we will see the change of shock profile due to different supra-thermal coefficient $(\kappa)$. For the existence of shock or solitary profile the value of this supra-thermal coefficient $\kappa$ must be greater than $\frac{3}{2}$, that is why we have taken all the values of $\kappa$ greater than this value. Also, when we are taking the values of $\kappa$ from $2.5-7.5$, we keep in mind the Ulysses, Cluster and Helioes observation for slow SW $e^-$ \cite{livadiotis2015statistical,2009JGRA..114.5104S}. In this case, the dispersive and dissipative coefficient of the shock are remained constant, the only changes can happen in the nonlinear coefficient and this thing is also clear from eq. \ref{eq33}. It is well known that  the Kappa parameter is measure of the discreteness of plasma concentration in space plasma, that is why increase of kappa parameter influences the greater shock. Two fluids can produce shock by propagating along side or colliding to each other. There are two kinds of shock that can form in case of fluids the first and mostly found in a viscous plasma is the strong shock when two fluids are propagating or colliding with each other with a velocity greater than the local acoustic waves, i.e. $(M>1)$ and the second kind of shock when two fluids collide a subsonic speed $(M<1)$ which is known as weak shock, this type of shock is very well known in case of fluids \cite{Zel'dovich,landau1959fluid} and for laser plasma interaction \cite{goswami2020collision} but till now no report has been made in the case of dusty plasma. In figures \ref{transitionfromweaktostrongshock}, we have discussed the transition of weak to strong shock and in figure \ref{Weakandstrongshock}, we have shown two different kinds of shock for subsonic $(M=0.7)$ and supersonic $(M=1.2)$ frame velocity.
\par
\begin{figure}
    \centering
     \includegraphics[width=10cm]{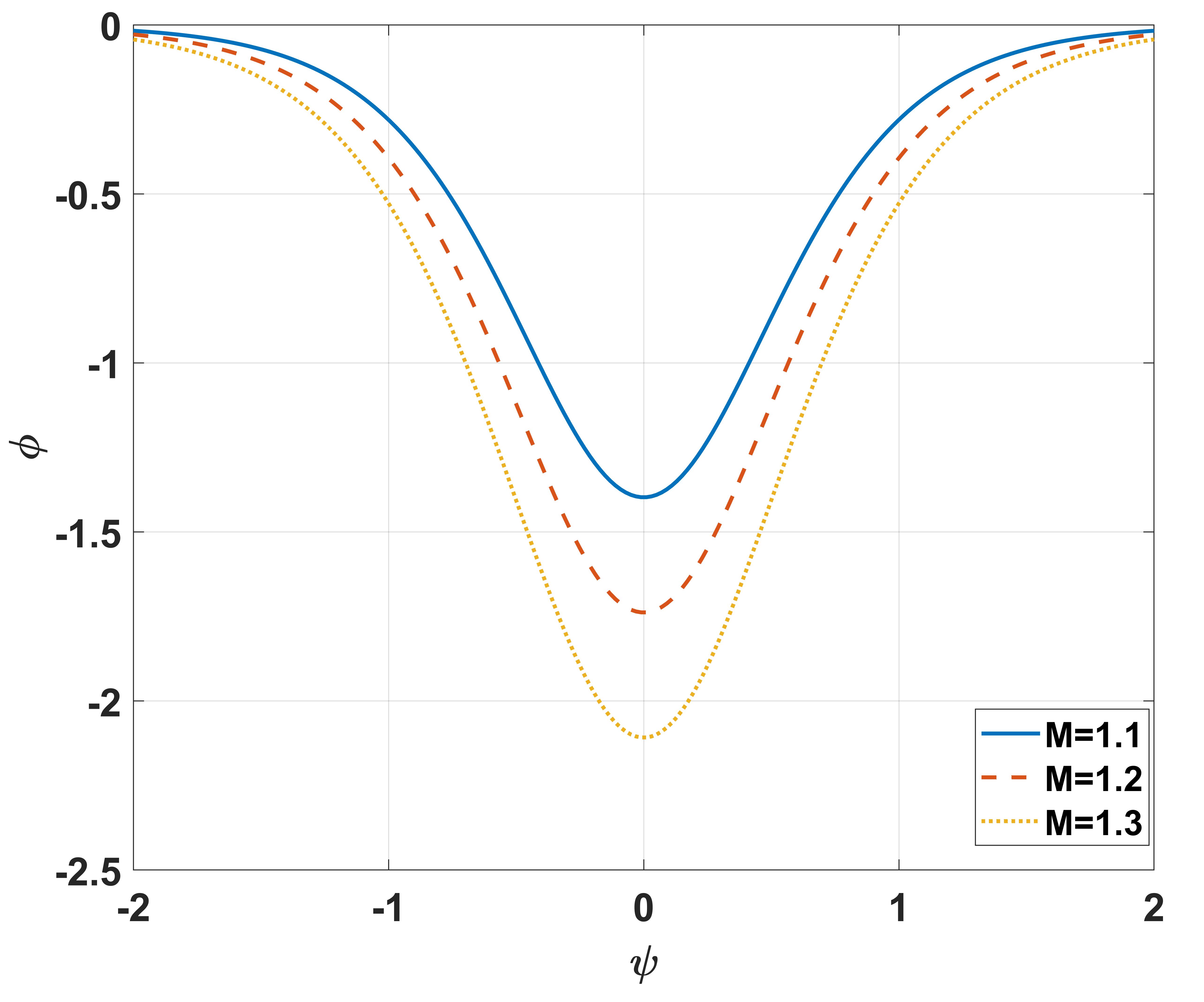}
    \caption{Solitary for different Mach numbers $(M)$ with unperturbed ion streaming velocity$(u_{i0})=0.3$, unperturbed dust streaming velocity$(u_{i0})=0.07$, finite temperature coefficient $(\sigma)=0.65$, kappa index $(\kappa)=5$ and dust charge $(q)=1$}
    \label{solitaryfordifferentmachnumber}
\end{figure}
\begin{figure}
    \centering
     \includegraphics[width=10cm]{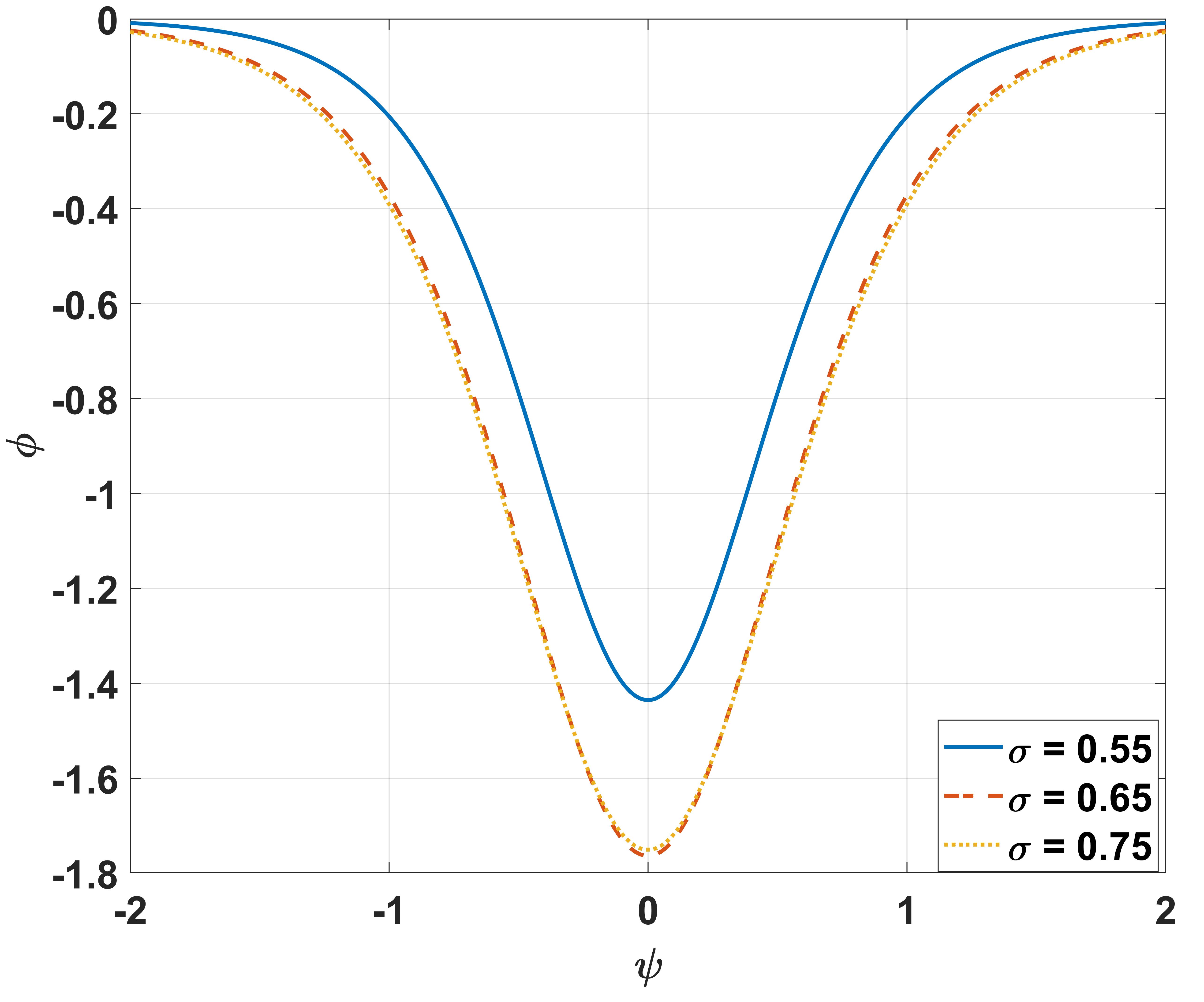}
    \caption{Solitary for different finite temperature coefficients $(\sigma)$ with unperturbed ion streaming velocity$(u_{i0})=0.3$, unperturbed dust streaming velocity$(u_{i0})=0.07$, Mach number $(M)=1.2$, kappa index $(\kappa)=5$ and dust charge $(q)=1$}
    \label{solitaryfordifferentsigma}
\end{figure}
\begin{figure}
    \centering
     \includegraphics[width=10cm]{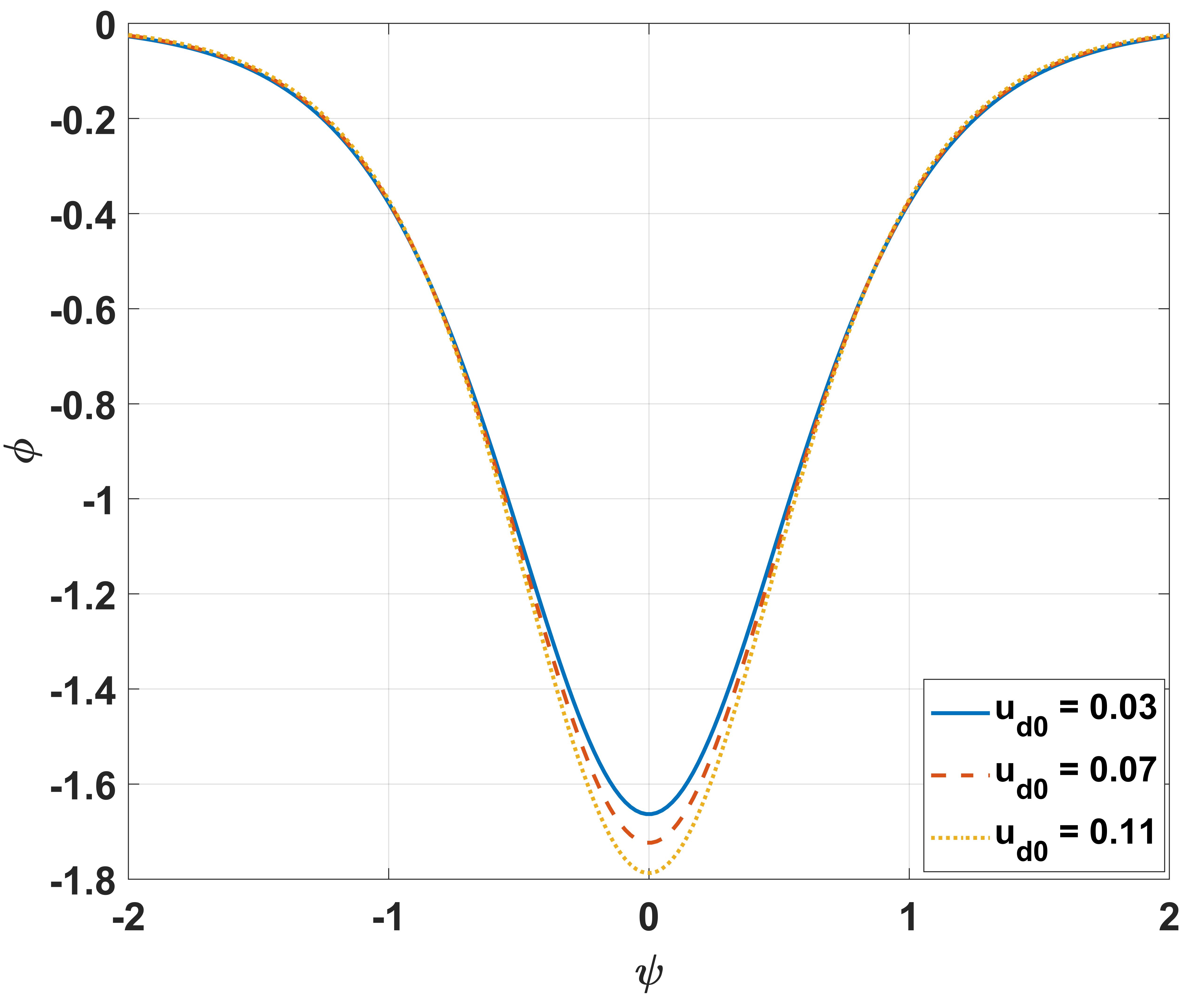}
    \caption{Solitary for different unperturbed dust streaming velocities $(u_{d0})$  with unperturbed ion streaming velocity$(u_{i0})=0.3$, finite temperature coefficient $(\sigma)=0.65$, Mach number $(M)=1.2$, kappa index $(\kappa)=5$ and dust charge $(q)=1$}
    \label{solitaryfordifferentstreamingofdust}
\end{figure}

Now as we have discussed in the section \ref{sec4B} when there is no viscous term the KdVB transforms into KdV equation and here we will discuss the analytical solution of KdV equation, i.e the solitary waves. In figure \ref{solitaryfordifferentmachnumber}, we can see that with an increase in the Mach Number $(M)$, the nonlinearity and dispersive coefficient decrease and increase, respectively. This has the same effect as we have seen in case of the figure \ref{shockfordifferentmachnumber}. In this case and also in the figure \ref{shockfordifferentmachnumber}, the Mach number $(M)$ also has a critical value i.e. $M_c=1.3$. Greater than this value $(M>M_c)$ the polarity of the shock waves as well as the solitary waves will change, that means all these rarefactive shock and solitary wave structures become compressive beyond this value of $M$. In the figure \ref{solitaryfordifferentsigma}, we have discussed the variation of solitary profile for different finite temperature degeneracy parameters $(\sigma)$. Initially, the nonlinear coefficient decreases and dispersive \& dissipation coefficient increases with the increase of the finite temperature degeneracy parameter. When the degeneracy parameter $\sigma$ increases greater than $0.661$ the nonlinear coefficient starts to increase along with other two parameters and that is why there is a different tendency of the solitary waves and this is same for cases of the shock waves too as shown in the figure \ref{shockfordifferentsigma}. Now, in the figure \ref{solitaryfordifferentstreamingofdust}, we see that the streaming velocity of dust cannot alter the width but increases the amplitude of the solitary wave. This has happened due to simultaneous decease of the nonlinear and dispersive coefficient of the solitary wave with the increase of dust streaming velocity. 
\begin{figure}
    \centering
     \includegraphics[width=10cm]{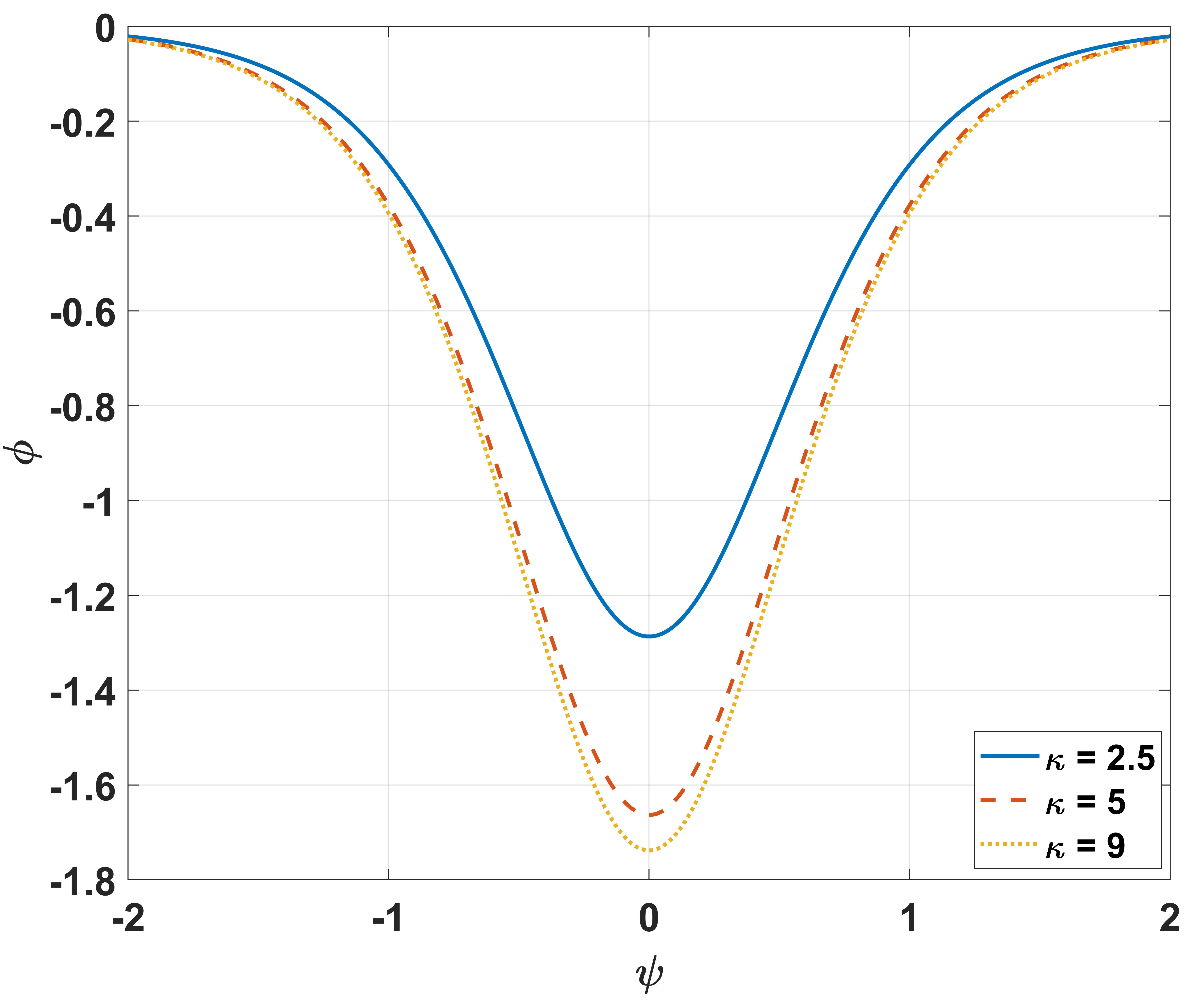}
    \caption{Solitary for different kappa indices $(\kappa)$  with unperturbed ion streaming velocity$(u_{i0})=0.3$, unperturbed dust streaming velocity$(u_{d0})=0.07$, finite temperature coefficient $(\sigma)=0.65$, Mach number $(M)=1.2$, and dust charge $(q)=1$}
    \label{solitaryfordifferentkappa}
\end{figure}
\begin{figure}
    \centering
     \includegraphics[width=10cm]{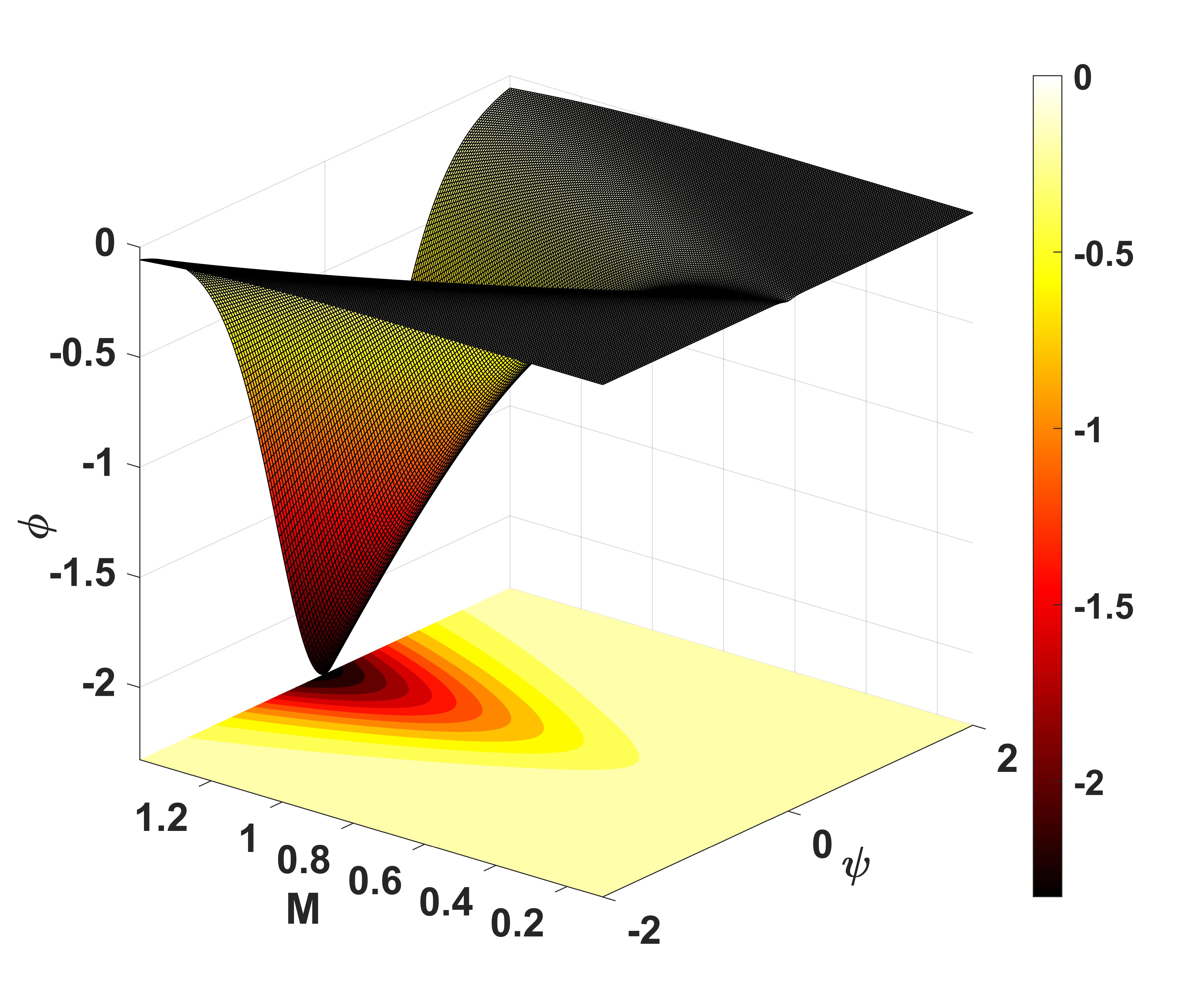}
    \caption{Solitary $3D$ for different Mach Numbers $(M)$ with unperturbed ion streaming velocity$(u_{i0})=0.3$, unperturbed dust streaming velocity$(u_{d0})=0.07$, finite temperature coefficient $(\sigma)=0.65$, kappa index $(\kappa)=5$ and dust charge $(q)=1$}
    \label{solitary3dfordifferentmachnumber}
\end{figure}
\begin{figure}
    \centering
     \includegraphics[width=10cm]{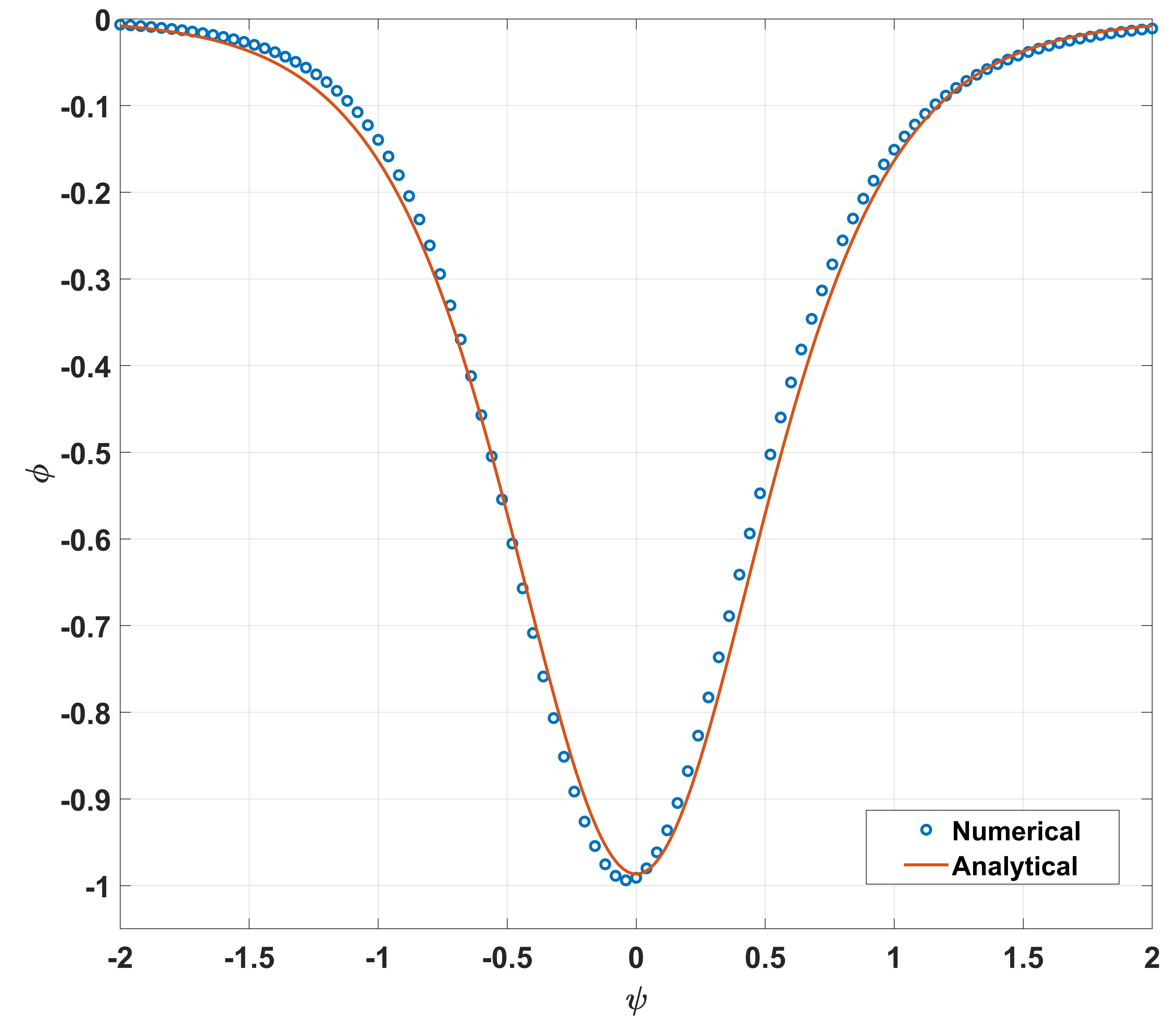}
    \caption{Numerical and analytical KdV}
    \label{Numericalanalyticalkdv}
\end{figure}
\begin{figure}
    \centering
     \includegraphics[width=10cm]{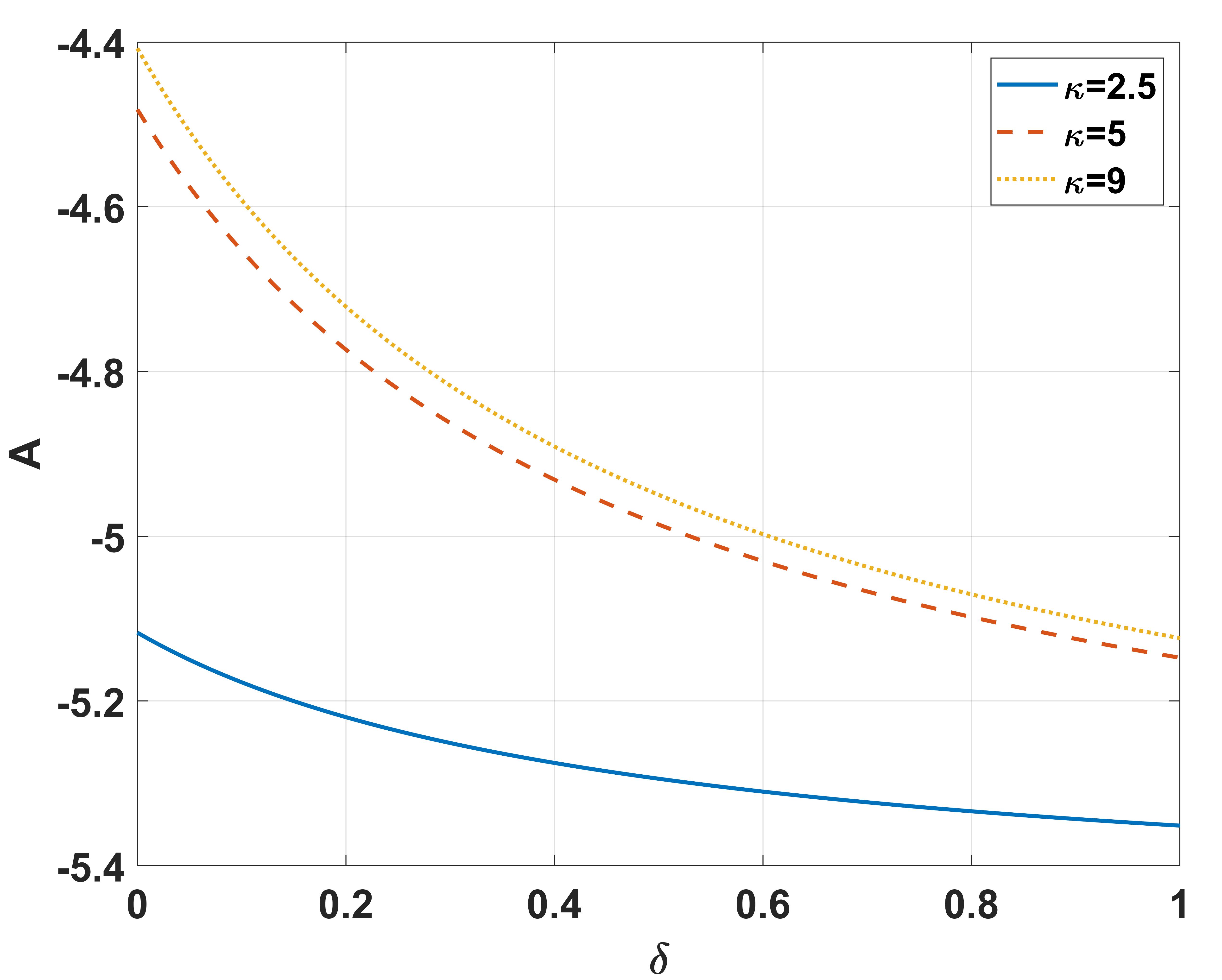}
    \caption{Non-linear coefficient for different Kappa $(\kappa)$}
    \label{nonlinearcoefficientfordifferentkappa}
\end{figure}
\begin{figure}
    \centering
     \includegraphics[width=10cm]{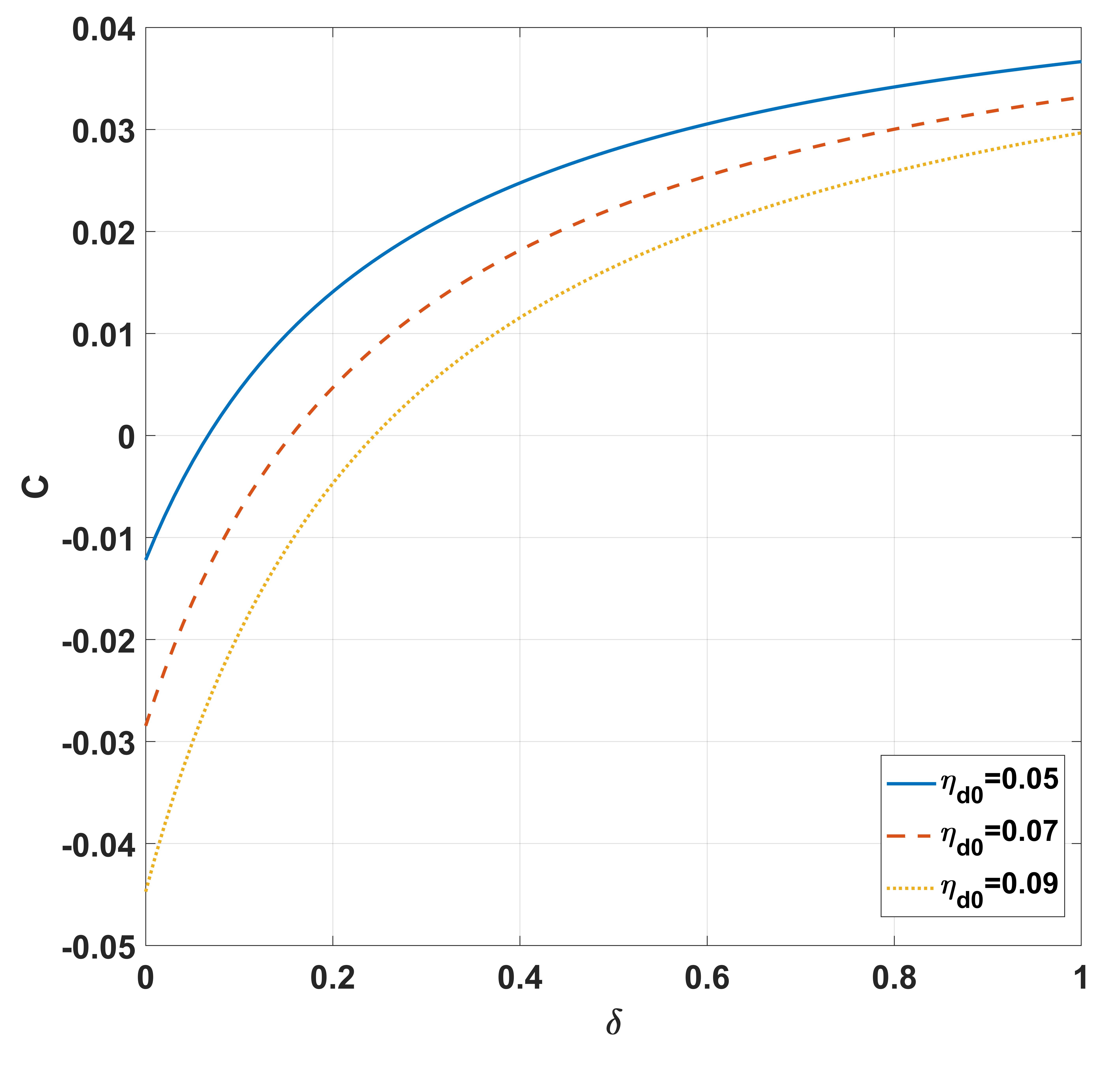}
    \caption{Dissipative coefficient for different viscous coefficient of dust $(\eta_{d0})$}
    \label{dissipativecoefficientfordiffeenteta}
\end{figure}
\par
In figure \ref{solitaryfordifferentkappa}, we have discussed the formation and variation of solitary profile for various supra-thermal parameters $(\kappa)$. From equation \ref{eq33}, it is very clear that the only coefficient that is affected by supra-thermal parameter $(\kappa)$ is the nonlinear coefficient $A$ and it has a decreasing tendency with the increasing value of $\kappa$ that is why the width and amplitude is increasing with the increased value of $\kappa$. Next we have shown in the figure \ref{solitary3dfordifferentmachnumber} the variation of solitary profile due to different Mach numbers that is the transition of the solitary profile starting from the subsonic velocity region and ending in the supersonic velocity region for the wave. In the figure \ref{Numericalanalyticalkdv}, we have plotted the exact solution of KdV from equation \ref{eq37} and the numerical solution of the same with the help of set of equation from \ref{eq38}- \ref{eq44}. For the numerical solution, we have considered the space like steps as $\Delta \xi=0.04$ and time like steps as $\Delta\tau=0.05$, the maximum error in $\phi^{(1)}$ from the exact to numerical solution is $0.009152$.
\par
At last we will discuss the variation of non-linear coefficient with ion density $(\delta)$ for different supra-thermal coefficient $(\kappa)$ (\ref{nonlinearcoefficientfordifferentkappa}) and dissipative coefficient with ion density $(\delta)$ for different viscous coefficients of dust $(\eta_{d0})$ (\ref{dissipativecoefficientfordiffeenteta}). It is very much evident from equation \ref{eq33} that two kinds of shock/ soliton may be possible to form. One is compressive  shock/ soliton and another rarefactive shock/ soliton structures in this scenario and for that the nonlinear coefficient $A>0$ and $A<0$, respectively. But from figure \ref{nonlinearcoefficientfordifferentkappa}, we can see the nonlinear coefficient is throughout negative  that is why we are getting rarefactive shocks and solitons. Again as $\phi_0 = \frac{3M}{A}$ so due to increasing value of $A$ will cause the decrease in the amplitude. So, this variation of $A$ with different $\kappa$  is also supported by the figures \ref{shockfordifferentkappa} and \ref{solitaryfordifferentkappa}.
\begin{figure}
    \centering
     \includegraphics[width=10cm]{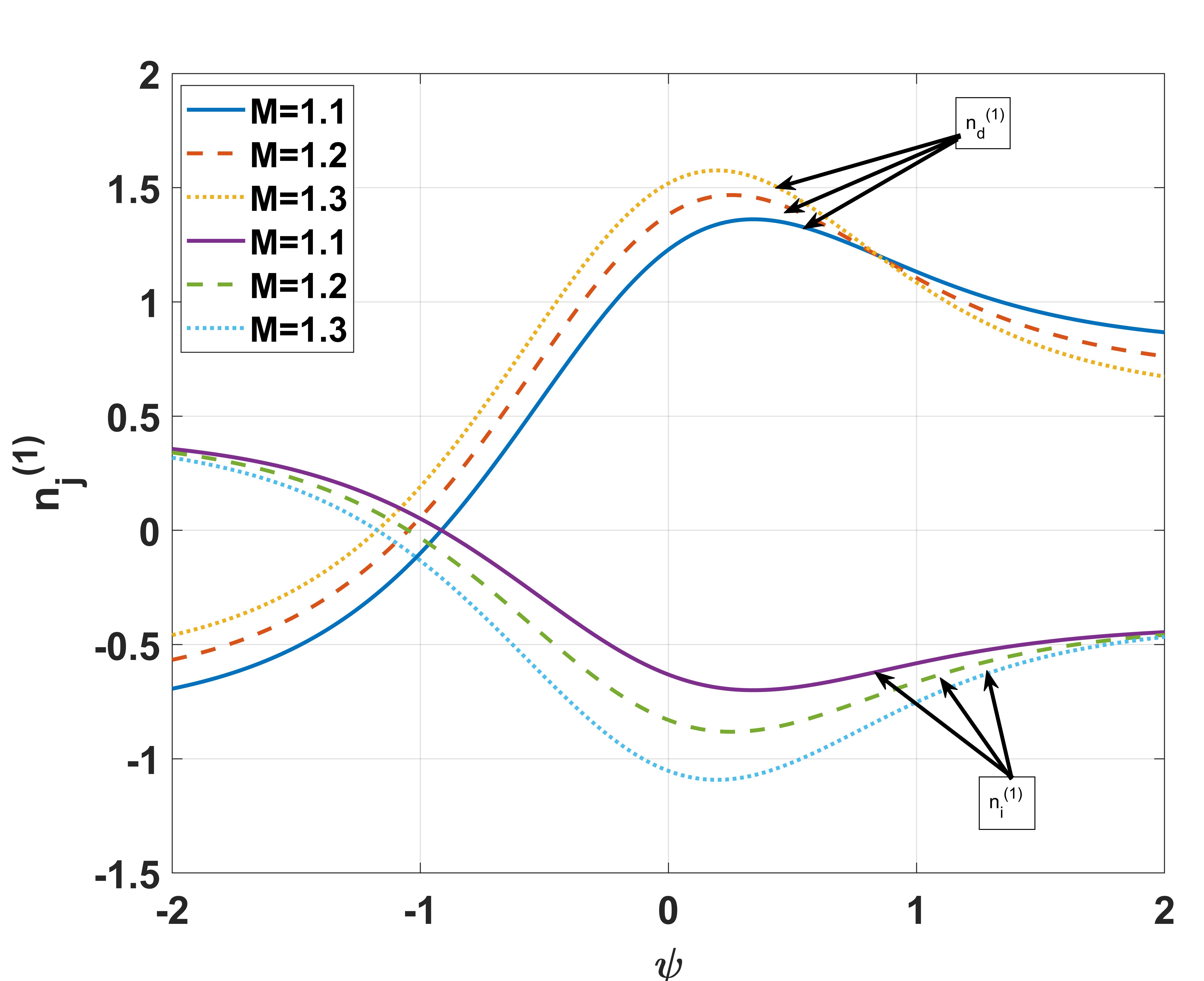}
    \caption{First order perturbed density of dust and ion with viscosity}
    \label{firstorderperturbeddensityofdustandionwithviscosity}
\end{figure}
\begin{figure}
    \centering
     \includegraphics[width=10cm]{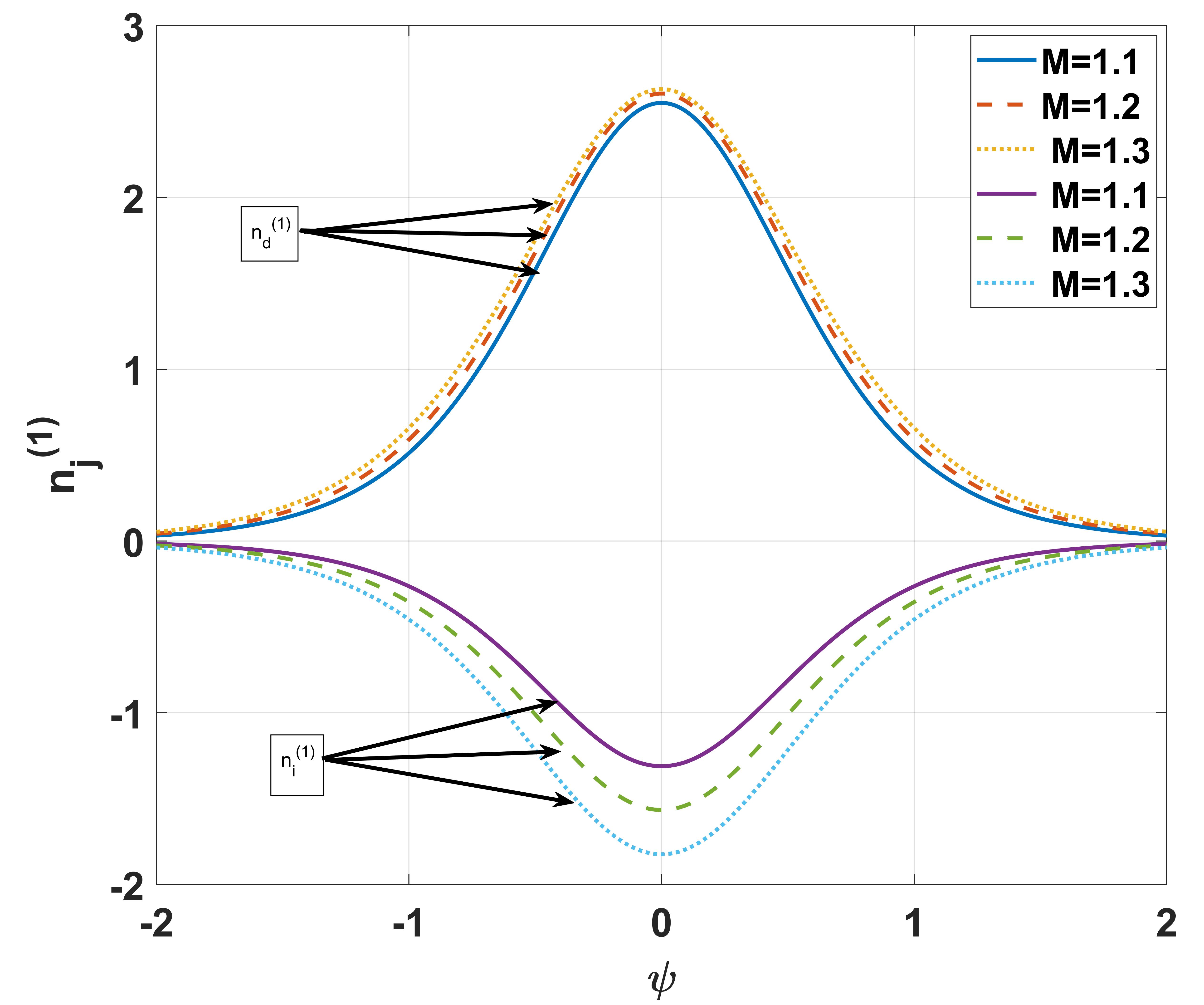}
    \caption{First order perturbed density of dust and ion without viscosity}
    \label{firstorderperturbeddensityofdustandion}
\end{figure}

In figures \ref{firstorderperturbeddensityofdustandionwithviscosity} and \ref{firstorderperturbeddensityofdustandion}, we have discussed the first order perturbed density change of dusts and ions in presence and absence of the viscosity with different Mach number $(M)$. When viscous force is absent in medium (figure \ref{firstorderperturbeddensityofdustandion}), we can see that the density of dust has a very little change with various $M$. This is due to heavy mass of the dust than ion, and also the density of dust and ion has different polarities. But from figure \ref{firstorderperturbeddensityofdustandionwithviscosity}, we can see that the effect of viscosity has larger impact than the effect of mass as in presence of viscosity there is a sufficient amount of changes in the dust density with the change in the Mach number $(M)$ as the viscosity considered in this paper is kinetic in nature.
\begin{figure}
    \centering
     \includegraphics[width=10cm]{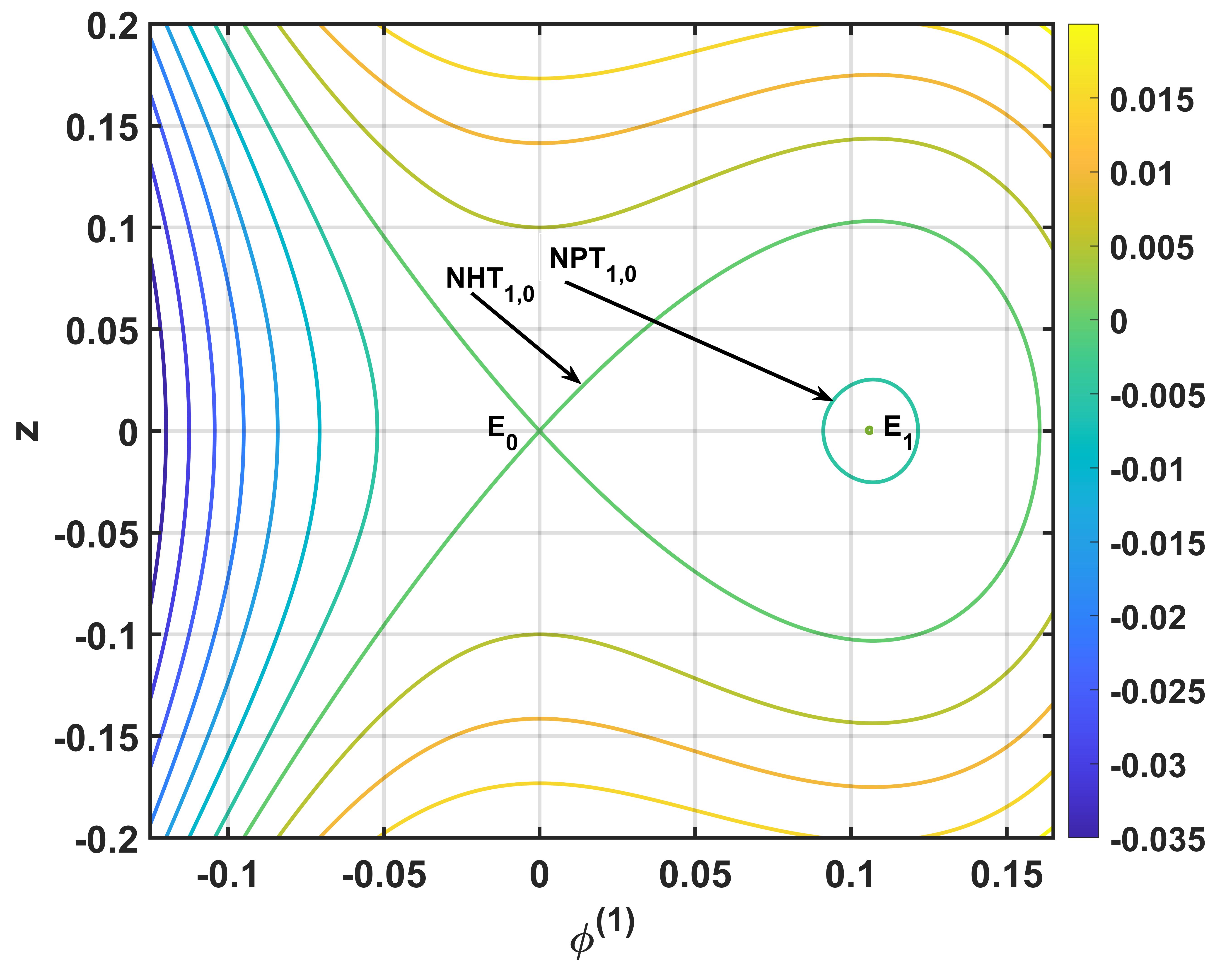}
    \caption{Phase plot of the system (\ref{eq45}) with unperturbed ion streaming velocity$(u_{i0})=0.3$, finite temperature coefficient $(\sigma)=0.65$, Mach number $(M)=1.2$, kappa index $(\kappa)=5$ and dust charge $(q)=1$}
    \label{phasediagramkdv}
\end{figure}

\par
The Hamiltonian $H(\phi,z)$ is given in equation \ref{eq49} defines the trajectory of the phase plot is given in figure \ref{phasediagramkdv}. In this figure \ref{phasediagramkdv}, we can easily define two equilibrium points $E_0(0,0)$ and $E_1(\frac{v}{A},0)$ of the system equation \ref{eq45} as saddle point and center. The solitary wave solution \ref{eq37} of equation \ref{eq36} represents the homoclinic orbit of the system \ref{eq45}. And periodic travelling wave solution of equation \ref{eq36} represents the periodic orbit of the system \ref{eq45}. In figure \ref{phasediagramkdv}, the phase portrait of equation \ref{eq45} is plotted. With the presence of the plasma parameters like $M=1$, $u_{i0}=0.3$, $u_{d0}=0.07$, $\eta_{i0}=0.9$, $\eta_{d0}=0.07$, $\sigma=0.7$ and $\kappa=5$, we have shown the saddle point $E_0(0,0)$ with no separatrix enclosed by the nonlinear homoclinic trajectory $(NHT_{T,0})$ which is obviously corresponds to the solitary wave solution shown in equation \ref{eq36}. Again around the critical point $E_1(\frac{v}{A},0)$ corresponding to the periodic solution there exists the family of nonlinear periodic trajectories $(NPT_{1,0})$ \cite{dubinov2012ion,Shome2021}. The corresponding analytical solution of nonlinear homoclinic trajectory $(NHT_{T,0})$ have already been shown in the figures \ref{solitaryfordifferentmachnumber}-\ref{solitary3dfordifferentmachnumber}.

\section{Conclusion:}\label{sec7}
In the light of the calculation and study this paper is divided into three parts, firstly with the help of reductive perturbation technique (RPT) we analytically derived and solved the KdV-Burger and KdV equation and get the shock and solitary waves as there solution, secondly using finite difference method we have done the numerical solution of KdV and lastly with the help of total energy of the system when there is no dissipative force present in the system we have done the phase plane analysis.
\par
In the model section we have discussed three things very broadly
\par
\space(a) The presence of Fermi-Dirac statistics in case of ion distribution. Various astrophysical and laboratory plasmas are in this category if they follow the condition we have mentioned in the section \ref{sec2B}. $H^+$ plasma is one of this kind of plasma. 
\par
\space(b) The detailed discussion of kinematic viscosity in both dust and ion species.
\par
\space(c) The reason behind taking the dust charge as constant in DIA case has also been discussed in section \ref{sec2C}.
\par
In the production of shock waves the dust kinematic viscous coefficient $(\eta_{d0})$ is the most important parameter, so we are very careful in taking the value of this. From the work of Vorona \emph{et al.} \cite{vorona2007viscosity} we can have some ideas what value should be for this parameter and tried to follow this in the manuscript. For the figures \ref{shockfordifferentmachnumber}-\ref{shockfordifferentstreamingofdust}, \ref{shockfordifferentkappa}-\ref{Weakandstrongshock} and \ref{firstorderperturbeddensityofdustandionwithviscosity} we have taken the kinematic viscosity of dust $n_{d0}=0.07$ and for figures \ref{shockfordifferentetad} and  \ref{dissipativecoefficientfordiffeenteta} it is shown in the legends. 
\par
Keeping the above mentioned values of viscous coefficient and other physically admissible values for different plasma parameters for the numerical study we have found some of the important results that we have mentioned below.
\par
\space(a) For the first time we have reported two types of shock that is weak and strong shock in a DIA mode.
\par
\space(b) Also physically accepted values for the super sonic wave region is also discussed.
\par
\space(c) The critical value of finite temperature degeneracy parameter $\sigma_c$ have also been discussed and again what values would be physically acceptable for the supersonic wave region have been discussed.
\par
\space(d) The critical value of the Mach number is also discussed in this result beyond which the polarity of shock and solitary waves would changes i.e. rarefactive shock/soliton transform in compressive shock/soliton.
\par
If we want to verify in accordance with the result in laboratory environment we have made some changes, one of these changes is that we have to take Boltzmann distribution of electron and it can easily happened by taking very large value of $\kappa$, as we know when $\kappa \xrightarrow[]{} \infty$ the Kappa distribution transforms into Boltzmann distribution. 

\bmhead{Acknowledgments}Authors would like to thank Prof. B. K. Saikia, Acting Centre Director, Centre of Plasma Physics – Institute for Plasma Research for providing facility and environment to produce this research article. One of the authors JG extend his heartfelt thanks to Gunjan Sharma, Rupali Paul and Kishor Deka for their valuable suggestions to improve the quality of the article. 

\bibliography{sn-bibliography}


\end{document}